\documentclass[hidelinks,onefignum,onetabnum]{siamart220329}

\usepackage{lipsum}
\usepackage{amsfonts}
\usepackage{graphicx}
\usepackage{comment}
\usepackage{epstopdf}
\usepackage{algorithmic}
\usepackage{tikz}
\usetikzlibrary{shapes, positioning, calc}
\ifpdf
  \DeclareGraphicsExtensions{.eps,.pdf,.png,.jpg}
\else
  \DeclareGraphicsExtensions{.eps}
\fi

\newsiamremark{Remark}{Remark}
\newsiamremark{Proof}{Proof}
\newsiamremark{hypothesis}{Hypothesis}
\crefname{hypothesis}{Hypothesis}{Hypotheses}
\newsiamthm{claim}{Claim}

\newsiamthm{Theorem}{Theorem}
\newsiamthm{Lemma}{Lemma}
\newsiamthm{Corollary}{Corollary}
\newsiamthm{Assumption}{Assumption}
\newsiamthm{Definition}{Definition}
\headers{Biased stochastic gradients under local geometry}{Bowen Song et al.}

\title{Biased Stochastic Gradient Descent under Local Geometry, with Application to Data-Driven LQR }

\author{Bowen Song \thanks{Institute for Systems Theory and Automatic Control, University of Stuttgart, Germany 
  (\email{bowen.song@ist.uni-stuttgart.de, andrea.iannelli@ist.uni-stuttgart.de}).}
\and Simon Weissmann \thanks{Institute for Mathematics, University of Mannheim, Germany
  (\email{simon.weissmann@uni-mannheim.de}).}
\and Mathias Staudigl \thanks{Department of Business Analytics and Decision Sciences, University of Economics and Business, Austria (\email{mathias.staudigl@wu.ac.at}).}
\and Andrea Iannelli $^\dag$}

\usepackage{amsopn}

\ifpdf
\hypersetup{
  pdftitle={Biased Stochastic Gradient Descent under Local Geometry, with Application to Data-Driven LQR},
  pdfauthor={Bowen Song, et al}
}
\fi

\begin{document}

\maketitle

\begin{abstract}
We study stochastic gradient descent (SGD) with biased stochastic gradient oracles for objectives that are only locally smooth and locally gradient dominated, with constants that may grow with the objective value and are required only on a sublevel set. Such oracles arise in data-driven optimization and control, where the gradient must be estimated from noisy data through nonlinear procedures such as model identification or smoothing and truncation. We show that if the oracle bias decays sufficiently fast, even only in expectation, and the second moment of the oracle remains bounded, then, on an event of prescribed probability, the optimality gap converges to zero almost surely with an asymptotic rate, and in expectation with an explicit bound at every iteration that separates the contributions of the initial condition, the oracle bias, and the gradient noise. We illustrate the framework on the linear quadratic regulator (LQR), whose cost satisfies these properties only on sublevel sets and whose policy gradient depends on the unknown system matrices. We consider an indirect estimator, which identifies the system matrices online, and a direct zeroth-order estimator. We characterize the bias and the second moment of both estimators and show how to choose their parameters so that the convergence conditions hold. On an event of prescribed probability, both data-driven policy gradient methods converge to the optimal policy almost surely and in expectation, with guarantees for all iterations rather than at a single prescribed iteration. We compare the two estimators, and numerical experiments on the LQR illustrate the results.
\end{abstract}



\section{Introduction}
   Reinforcement learning (RL) \cite{sutton2018reinforcement} has had a profound impact across a wide range of applications \cite{annurev:/content/journals/10.1146/annurev-control-053018-023825,annurev:/content/journals/10.1146/annurev-control-042920-020021}.
 A central component of RL is policy optimization, in which a parameterized policy is directly optimized with respect to a prescribed performance objective \cite{annurev:/content/journals/10.1146/annurev-control-042920-020021}, most commonly by policy gradient (PG) methods. Since the gradient of the performance objective is rarely available exactly, PG methods are in practice implemented as stochastic gradient descent (SGD) \cite{Robbins1951ASA,spall2003introduction,doi:10.1137/16M1080173,garrigos2024handbookconvergencetheoremsstochastic} with gradients estimated from sampled data. Understanding when such methods converge to the optimal policy in the presence of uncertainty and stochastic disturbances is essential for their reliable deployment \cite{JMLR:v22:19-736,pmlr-v162-wang22at}, and raises a general question about the convergence of SGD for objectives and gradient estimators that violate the classical assumptions.

\subsection*{Data-driven policy gradient for the LQR}
The linear quadratic regulator (LQR) is a canonical benchmark for RL in continuous state and action spaces \cite{pmlr-v80-fazel18a,annurev:/content/journals/10.1146/annurev-control-053018-023825}. Its cost is gradient dominated and smooth on every sublevel set, but infinite outside the set of stabilizing gains. Since \cite{pmlr-v80-fazel18a} established global convergence of PG methods for the LQR with exact gradients, this setting has been studied extensively, see \cite{annurev:/content/journals/10.1146/annurev-control-053018-023825,doi:10.1137/20M1347942} and the references therein, together with the robustness of PG to inexact gradients \cite{cui2025perturbedgradientdescentalgorithms,9444823}. When the system is unknown, the gradient must be estimated from data. Indirect methods identify the system matrices and evaluate the gradient at the estimated model \cite{SongIannelli+2025+398+412,zhao2025policygradientadaptivecontrol}, whereas direct methods bypass identification, either by estimating the quantities required for the update \cite{doi:10.1137/23M1554771} or through zeroth-order gradient estimates \cite{pmlr-v80-fazel18a,Full,doi:10.1137/20M1382386}. These methods are typically analyzed by bounding the gradient estimation error at each iteration with high probability and combining it with the per-iteration decrease of the cost. The resulting union bound over the iterations is conservative: it requires many samples per iteration and a confidence level that degrades with the number of iterations. In~\cite{SongIannelli+2025+398+412,zhao2025policygradientadaptivecontrol}, the noise is treated as a bounded worst-case sequence, in~\cite{zhao2025policygradientadaptivecontrol} subject to a lower bound on the signal-to-noise ratio. The analyses closest to an SGD viewpoint are~\cite{pmlr-v89-malik19a,JMLR:v26:24-0636,li2025robustnessderivativefreemethodslinear}, which study direct methods with a cost oracle; their guarantees state that a prescribed accuracy $\varepsilon$ is reached at an $\varepsilon$-dependent iteration with a fixed probability, as in~\cite{talebi2023datadriven} for the dual filtering problem. Each of these analyses is tailored to a specific algorithm and estimator, and none provides guarantees for all iterations, almost surely, or for estimators that reuse past data under stochastic noise.

\subsection*{Stochastic gradient descent under local conditions and biased oracles}
Stochastic gradient methods go back to \cite{Robbins1951ASA}; see \cite{doi:10.1137/16M1080173,garrigos2024handbookconvergencetheoremsstochastic} for an overview. The classical theory assumes a globally $L$-smooth objective and an unbiased stochastic gradient with bounded variance. Under strong convexity, both asymptotic and non-asymptotic rates are well established \cite{polyak1987introduction,NIPS2011_40008b9a,Nguyen2018NewCA}. Without convexity, SGD is in general only guaranteed to converge to stationary points \cite{doi:10.1137/120880811,3666122.3667871}; under gradient domination, or the Polyak--{\L}ojasiewicz condition \cite{POLYAK1963864,10.1007/978-3-319-46128-1_50}, global optimality is retained \cite{bassily2018exponentialconvergencesgdnonconvex,Liu2020LossLA,NEURIPS2022_65ae674d,khaled2022better}, see also \cite{pmlr-v134-fontaine21a,masiha2024complexityminimizingprojectedgradientdominatedfunctions} for related generalizations. These results assume smoothness and gradient domination on the whole parameter space, which fails for the LQR. Several works relax smoothness: under $(L_0,L_1)$-smoothness \cite{vankov2025optimizingl0l1smoothfunctions} and its generalization \cite{3666122.3667871}, the smoothness constant depends on the gradient norm, but gradient domination is not exploited. Others impose conditions only near the iterates or a minimizer and conclude on the event that the iterates remain there: \cite{NEURIPS2020_0cb5ebb1} assumes global smoothness and studies convergence near local minimizers, \cite{JMLR:v21:19-636} assumes a non-degenerate Hessian along the set of minima, \cite{Dereich2024} proves almost sure convergence to critical points under a {\L}ojasiewicz inequality on the event that the iterates stay bounded, \cite{pmlr-v162-scaman22a} derives a high-probability bound for the best iterate under a {\L}ojasiewicz condition with a fixed smoothness constant on a ball around the initialization, and \cite{Almost} establishes almost sure rates on the event that the iterates remain in a sublevel set on which gradient domination holds, but with a global smoothness constant and an unbiased oracle. 

All these analyses assume an unbiased oracle. This fails when the gradient is constructed from data through a nonlinear procedure, such as evaluating the gradient formula at an estimated model, or approximating the gradient from noisy function evaluations as in zeroth-order methods \cite{doi:10.1137/120880811,NEURIPS2018_ba9a56ce,10.1007/s10208-021-09499-8,OPT-047}; a bias is also introduced by algorithmic modifications such as clipping \cite{Mukherjee2026}. Biased oracles are considered in the inexact-oracle framework of \cite{Devolder2014} for deterministic convex optimization, where gradient methods reach the optimum only up to the oracle accuracy (see also~\cite{ajalloeian2020convergence,demidovich2023guide} for SGD under the Polyak--{\L}ojasiewicz condition), and in \cite{9736646}, which derives non-asymptotic bounds for SGD with globally smooth objectives and a bias controlled by the oracle parameters. Closest to the present setting, \cite{10.1007/s10957-024-02547-7} establishes almost sure convergence and rates for SGD with a biased oracle under gradient domination, but requires smoothness, gradient domination and the oracle bounds on the whole space, and a conditional bias bounded almost surely by a vanishing deterministic sequence. Neither requirement holds for data-driven PG for the LQR: the cost is infinite outside the stabilizing gains, all its properties hold only on sublevel sets, and an estimator that reuses past data has a random bias that only decays in expectation. To our knowledge, SGD has not been analyzed when smoothness, gradient domination and the oracle bounds all hold only locally, and the bias vanishes only in expectation. 

\subsection*{Contributions}
We analyze SGD with a biased stochastic gradient oracle for objectives that are only locally smooth and locally gradient dominated, with constants that may grow with the objective value and are required only on a sublevel set, and with an oracle whose bias is random and vanishes only in expectation. We show that the iterates remain in a sublevel set with a prescribed probability, using a confinement argument that builds on \cite{Almost}, and that, on this event, the optimality gap vanishes almost surely with an asymptotic rate, while its expectation satisfies an explicit bound at every iteration. This bound separates the contributions of the initial condition, the bias and the gradient noise. The step sizes do not depend on a target accuracy, so that the guarantees hold for all iterations rather than at a prescribed horizon. Compared with \cite{10.1007/s10957-024-02547-7}, which obtains almost sure rates of the same type, all conditions are local and the bias only needs to vanish in expectation.

We then apply this framework to data-driven PG for the LQR, with an indirect estimator based on recursive least squares, operated off-policy or on-policy, and a direct zeroth-order estimator. The indirect estimator reuses all past data, so that its bias is random and correlated across iterations, whereas the direct estimator draws fresh rollouts, so that its bias is deterministic given the current gain but its noise is unbounded. We characterize the bias and the second moment of both estimators and choose their parameters so that the conditions of the framework hold, which yields, on an event of high probability, convergence to the optimal policy almost surely and in expectation, with guarantees for all iterations. These guarantees are of a different type than those of existing SGD-based analyses of PG for the LQR~\cite{pmlr-v89-malik19a,JMLR:v26:24-0636,li2025robustnessderivativefreemethodslinear} and its dual filtering problem \cite{talebi2023datadriven}, which hold at a single, accuracy-dependent iteration with a fixed probability.
Finally, we compare the two estimators in terms of sample complexity, computation and excitation requirements.

Lemmas~\ref{bound}--\ref{quasi} and~\ref{GDDI} build on \cite{pmlr-v80-fazel18a,Full}, and Lemmas~\ref{lem:grad-error} and~\ref{lem:ms-bound} on
\cite{zhao2025policygradientadaptivecontrol}; all other results are new. The proof of Theorem~\ref{thm:main-convergence} is given in the main text;
all other proofs are deferred to the Appendix.

\subsection*{Notations}

In Sections~\ref{sec:abstract-setting}--\ref{sec:Convergence}, $\mathcal X$ denotes a finite-dimensional real inner product space with inner product $\langle\cdot,\cdot\rangle$ and Euclidean norm $\|\cdot\|$; from Section~\ref{sec:LQR-example} on, $\|\cdot\|$ denotes the induced 2-norm of a matrix, and $\|\cdot\|_F$ its Frobenius norm. We denote by $A\succeq 0$ and $A\succ0$ a positive semidefinite and a positive definite matrix $A$, respectively. $\mathbb{Z}_+$ and $\mathbb{Z}_{++}$ are the sets of non-negative and positive integers. 
A square matrix $A$ is Schur stable if $\rho(A)<1$, where $\rho(A)$ is its spectral radius, and $\lambda_i(A)$ denotes the $i$-th smallest eigenvalue of a symmetric matrix $A$. $I_n$ and $O_n$ are the identity and zero matrices of size $n\times n$. For $x\in\mathbb{R}$, $\lfloor x \rfloor$ denotes the greatest integer less than or equal to $x$, and $\lceil x \rceil$ the smallest integer greater than or equal to $x$. For a measurable set $\mathcal{A}$, the indicator function $\boldsymbol{1}_{\mathcal{A}}$ is defined by $\boldsymbol{1}_{\mathcal{A}}(w)=1$ if $w\in \mathcal{A}$ and $\boldsymbol{1}_{\mathcal{A}}(w)=0$ otherwise. A sequence is a map $\mathbb{Z}_+\rightarrow \mathbb{R}^{n \times m}$, denoted by $\{Y_t\}$, and its truncation up to index $N$ by $\{Y_t\}_{t=0}^{N}$. For scalar sequences $\{a_t\}$ and $\{g_t\}$ with $a_t\geq0$ and $g_t>0$, we write $a_t=O(g_t)$ if there exist $C>0$ and $t_0\in\mathbb{Z}_+$ such that $a_t\leq Cg_t$ for all $t\geq t_0$, and $a_t=o(g_t)$ if $\lim_{t\rightarrow+\infty}a_t/g_t=0$.

\section{Preliminaries}
\label{sec:abstract-setting}
We consider the minimization problem:
\begin{equation}
    \min_{X\in\mathcal{X}} f(X)
\end{equation}
where $\mathcal{X}$ denotes a finite-dimensional real inner product space with inner product $\langle\cdot,\cdot\rangle$ and Euclidean norm $\|\cdot\|$; the function $f:\mathcal{X} \rightarrow\mathbb{R}\cup\{+\infty\}$ denotes the objective function, which we assume to be continuously differentiable on $\operatorname{dom} f:=\{X\in\mathcal X: f(X)<+\infty\}$; Definitions~\ref{def:GD}–\ref{def:local-smooth} and Assumption~\ref{as:grad-bound} below are required only for $X\in\operatorname{dom} f$. Throughout this paper, we assume that the objective function attains its global minimum at $X^*$ with $f(X^*)\geq0$. Now we introduce the following definition of a sublevel set.
\begin{Definition}[Sublevel Set]
\label{def:level-set}
For $J \geq f(X^*)$, the sublevel set of $f$ at $J$ is defined as $\mathcal{S}(J) := \{X \in \mathcal{X} \mid f(X) \leq J\}.$
\end{Definition}

We assume that the function $f$ satisfies the following conditions, namely \emph{local gradient domination} \cite{POLYAK1963864} and \emph{local smoothness}:
\begin{Definition}[Local Gradient Domination]\label{def:GD}
The continuously differentiable function $f$ satisfies the local gradient domination property if there exists a function $\mu:\mathbb{R}_{\geq f(X^*)}\to\mathbb{R}_{>0}$ such that, for all $J\geq f(X^*)$ and all $X\in\mathcal{S}(J)$,
\begin{equation}\label{eq:GD}
    f(X)-f(X^*)\leq \mu(J)\,\|\nabla f(X)\|^2 .
\end{equation}
\end{Definition}
Classical gradient domination is recovered when $\mu$ is constant and $\operatorname{dom} f=\mathcal X$.
\begin{Definition}[Local Smoothness]\label{def:local-smooth}
    The continuously differentiable function $f$ satisfies the local smoothness property if there exist a non-decreasing function $L: \mathbb{R}_{\geq f(X^*)} \rightarrow \mathbb{R}_{>0}$ and a non-increasing function $r: \mathbb{R}_{\geq f(X^*)} \rightarrow \mathbb{R}_{>0}\cup \{+\infty\} $ such that, for all $X \in \operatorname{dom} f$ and $X' \in \mathcal X$ satisfying $\lVert X - X' \rVert \leq r(f(X))$, we have $X'\in\operatorname{dom} f$ and
    \begin{equation}
        \big| f(X') - f(X) - \langle \nabla f(X), X'-X \rangle
        \big| \leq \frac{L(f(X))}{2} \lVert X'-X \rVert^2 .
    \end{equation}
\end{Definition}
Classical global $L$-smoothness is recovered when $L$ is constant and $r\equiv\infty$, i.e., when neither the modulus nor the radius depends on the objective value. In general, global $L$-smoothness implies local smoothness, but not vice versa. In addition, we assume that the norm of the gradient $\nabla f$ is bounded by a function of the objective value.
\begin{Assumption}[Gradient Boundedness]
\label{as:grad-bound}
There exists a non-decreasing function $b_\nabla : \mathbb{R}_{\geq f(X^*)}
\rightarrow \mathbb{R}_{\geq 0}$ such that
\begin{equation}
    \|\nabla f(X)\| \leq b_\nabla\big(f(X)\big), \qquad \forall X \in
    \operatorname{dom} f.
    \label{eq:grad-bound}
\end{equation}
\end{Assumption}
This assumption is mild, as it holds whenever $f$ has compact sublevel sets. Together with Assumption~\ref{as:grad-bound}, the generalized smoothness of~\cite{3666122.3667871}, where $L$ and $r$ depend on $\|\nabla f(X)\|$ instead of $f(X)$, implies Definition~\ref{def:local-smooth}, with $L$ and $r$ evaluated at $b_\nabla(f(X))$; this includes $(L_0,L_1)$-smoothness~\cite{vankov2025optimizingl0l1smoothfunctions} as a special case.

\subsection{Stochastic Gradient Descent}
\label{subsec:sgd-algorithm}
We assume access to the gradient $\nabla f$ only through a stochastic first-order oracle, and consider the stochastic gradient descent (SGD) algorithm
\begin{equation}
    X_{i+1} = X_i - \eta_i \hat\nabla f(X_i, \xi_i), \qquad i \in
    \mathbb{Z}_+,
    \label{eq:sgd-iteration}
\end{equation}
initialized at some $X_0 \in \operatorname{dom} f$, where $\{\eta_i\}$ is a deterministic, positive step-size sequence, and $\hat\nabla f(X_i,\xi_i)\in\mathcal{X}$ is the oracle output at iteration $i$, with $\xi_i$ the randomness underlying the query, e.g., the noise in the trajectory data used to construct the estimate. The oracle may change across iterations, e.g., through a sample size, smoothing radius, or other internal parameter tuned as $i$ grows, and it may reuse information collected at previous iterations. The properties of the oracle are specified in Section~\ref{subsec:oracle}.

\section{Convergence Analysis of SGD with Biased Gradients}\label{sec:Convergence}
In this section, we first specify the oracle model (Section~\ref{subsec:oracle}) and then analyze the convergence of the SGD iteration \eqref{eq:sgd-iteration} to a minimizer $X^*$ of $f$ (Section~\ref{subsec:convergence}).

\subsection{Stochastic Biased Oracle}\label{subsec:oracle}
Throughout this section, we fix a level $J_0> f(X_0)$ and work on the sublevel set $\mathcal{S}(J_0)$ of Definition~\ref{def:level-set}. We write $\hat g_i:=\hat\nabla f(X_i,\xi_i)$ and let $\{\mathcal{F}_i\}$ be a filtration such that $X_0$ is $\mathcal{F}_0$-measurable and $\hat g_i$ is $\mathcal{F}_{i+1}$-measurable for all $i\in\mathbb{Z}_+$, e.g., $\mathcal{F}_i:=\sigma(X_0,\xi_0,\dots,\xi_{i-1})$; by \eqref{eq:sgd-iteration}, $X_i$ is then $\mathcal{F}_i$-measurable.
\begin{Assumption}[Biased Stochastic Gradient Oracle]
\label{as:oracle}
There exist constants $c(J_0)\geq0$, $\Delta_0(J_0)>0$ and $\beta>0$ such that, for all $i\in\mathbb{Z}_+$, almost surely on $\{X_i\in\mathcal{S}(J_0)\}$,
\begin{subequations}\label{eq:oracle}
\begin{align}
    \mathbb{E}\big[\hat g_i\mid\mathcal{F}_i\big] &= \nabla f(X_i)+\Delta_i, \label{eq:oracle-mean}\\
    \mathbb{E}\big[\|\hat g_i\|^2\mid\mathcal{F}_i\big] &\leq c(J_0), \label{eq:ABC}
\end{align}
\end{subequations}
where the bias $\Delta_i$ is $\mathcal{F}_i$-measurable and satisfies
\begin{equation}\label{eq:bias-decay}
    \mathbb{E}\big[\|\Delta_i\|\,\boldsymbol{1}_{\{X_i\in\mathcal{S}(J_0)\}}\big]\leq\Delta_0(J_0)\,(i+1)^{-\beta},\qquad\forall i\in\mathbb{Z}_+.
\end{equation}
\end{Assumption}
Conditions \eqref{eq:oracle-mean}--\eqref{eq:ABC} describe the oracle given the past: its conditional mean deviates from the gradient by the bias $\Delta_i$, and its conditional second moment is bounded. Condition \eqref{eq:bias-decay} requires the bias to vanish, but only in expectation over the whole history $X_0,\xi_0,\dots,\xi_{i-1}$. If the oracle draws fresh samples at each iteration, the bias depends on the past only through the current iterate, i.e., $\Delta_i=\Delta(X_i,i)$ for a deterministic function $\Delta$; if it reuses past data, the bias depends on the whole history. The notation $c(J_0)$ and $\Delta_0(J_0)$ emphasizes that the constants may depend on the chosen level $J_0$. Assumption~\ref{as:oracle} differs from the oracle models in the literature in three respects.

First, the bias is only required to decay in expectation. Classical analyses of SGD assume that the oracle error is zero-mean given the past~\cite{Almost,pmlr-v178-liu22d}, or that the conditional bias is bounded almost surely by a deterministic sequence~\cite{10.1007/s10957-024-02547-7} or by a deterministic function of the oracle parameters~\cite{9736646}; in the inexact-oracle framework of~\cite{Devolder2014}, the error is controlled at each iteration by a prescribed accuracy. These models fit oracles that draw fresh samples at each query, but not estimators that accumulate data over the iterations: given the past, the error of such an estimator is a random bias, correlated across iterations, which decays in expectation as data accumulate but admits no almost-sure bound by a deterministic sequence. Both cases arise in Section~\ref{sec:LQR-example}.

Second, both conditions are local: they are only required on a sublevel set of $f$, with constants that may grow with the objective value. In~\cite{10.1007/s10957-024-02547-7}, the conditional variance may grow linearly with the objective value and polynomially with the iteration index, but all bounds, including smoothness and gradient domination, are global; in~\cite{9736646}, smoothness, the gradient bound and the oracle bounds are global. Such conditions exclude estimators that are undefined or unbounded outside a region of interest. Local bounds suffice since the iterates remain in a sublevel set with high probability (Section~\ref{subsec:convergence}). Moreover, \eqref{eq:ABC} is implied by the condition $\mathbb{E}[\|\hat g_i\|^2\mid\mathcal{F}_i]\leq a_1(f(X_i)-f(X^*))+b_1\|\nabla f(X_i)\|^2+c_1$ of~\cite{khaled2022better,pmlr-v178-liu22d}, whose right-hand side is bounded on $\mathcal{S}(J_0)$ by Assumption~\ref{as:grad-bound}.

Third, the bias is treated as a design parameter that can be driven to zero, e.g., by collecting more data or by shrinking a smoothing radius, rather than as a perturbation of fixed size, under which gradient methods are in general only guaranteed to reach a neighborhood of the optimum whose size depends on the oracle accuracy~\cite{Devolder2014}.
\subsection{Convergence Analysis}\label{subsec:convergence}
Since the smoothness property of Definition~\ref{def:local-smooth} only holds within a ball of radius $r(f(X))$ around each point, we use that, on $\mathcal S(J_0)$, $r(f(X))\ge r(J_0)$ and $L(f(X))\le L(J_0)$ by monotonicity. The first step of the analysis is to show that, with high probability, the iterates remain in $\mathcal{S}(J_0)$ and consecutive iterates are within distance $r(J_0)$ of each other.

For $i\in\mathbb{Z}_+$, define the events
\begin{subequations}\label{eq:event-Fi}
    \begin{align}
        \Omega_i &:= \{X_k\in \mathcal{S}(J_0),\ k=0,\dots,i\},~ \mathrm{and}~\Omega := \bigcap_{i}\Omega_i,\\
        C_i &:= \{\|X_k - X_{k-1}\| \leq r(J_0),\ k=1,\dots,i\},~ \mathrm{and}~ C := \bigcap_{i} C_i,\\
        F_i &:= \Omega_i\cap C_i, ~ \mathrm{and}~F := \Omega\cap C = \bigcap_{i} F_i, \label{eventeqf}
    \end{align}
\end{subequations}
with $C_0$ the sure event. Each $F_i$ is $\mathcal{F}_i$-measurable and
$F_{i+1}\subseteq F_i$. Since the local properties of $f$ and of the oracle are only
available on $\mathcal{S}(J_0)$, confinement can only be lost at an iteration $k$ at which
$F_{k-1}$ holds, either through a step longer than $r(J_0)$ or through a short step
that leaves $\mathcal{S}(J_0)$. The next lemma controls the first case.

\begin{lemma}\label{lem:event-C}
Suppose that Assumption~\ref{as:oracle} holds. Consider $\{X_i\}$ generated by \eqref{eq:sgd-iteration} with positive step sizes $\{\eta_i\}$, initialized at $X_0$, and let
$J_0> f(X_0)$. Given $\delta\in(0,1)$, suppose the step sizes satisfy
\begin{equation}\label{eq:lemC-condition}
    \sum_{i=0}^\infty \eta_i^2 \leq \frac{r(J_0)^2\,\delta}{c(J_0)}.
\end{equation}
Then
\begin{equation}\label{eq:lemC-claim}
    \mathbb{P}\Big(\bigcup_{k\in\mathbb{Z}_{++}} F_{k-1}\cap\{\|X_k-X_{k-1}\|>r(J_0)\}\Big)\leq\delta.
\end{equation}
\end{lemma}

\begin{Lemma}
\label{lem:event-F}
Suppose that Definition \ref{def:local-smooth} and Assumptions~\ref{as:grad-bound} and \ref{as:oracle}  hold. Let $\{X_i\}$ be generated by \eqref{eq:sgd-iteration} with positive step sizes $\{\eta_i\}$, initialized at $X_0$ with $J_0 >f(X_0)$. Choose $\delta_1, \delta_2, \delta_3 \in (0,1)$ such that $\delta := \delta_1+\delta_2+\delta_3 < 1$, and define $\epsilon:=\big(\frac{\sqrt{1+4\epsilon'}-1}{2}\big)^2$ with $\epsilon':=J_0-f(X_0)>0$, so that $\sqrt{\epsilon}+\epsilon=\epsilon'$. Suppose that the step sizes satisfy
\begin{enumerate}
    \item $\displaystyle\sum_{i=0}^\infty \eta_i^2 \leq \frac{\delta_1 \epsilon}{\sigma_\zeta^2(J_0) + \frac{1}{2}L(J_0)\,c(J_0)}$ \ with $\sigma_\zeta^2(J_0):=b_{\nabla}(J_0)^2\big(\sqrt{c(J_0)}+b_\nabla(J_0)\big)^2$;\label{eq:step-cond-2}
    \item $\displaystyle\sum_{i=0}^\infty \eta_i\Delta_0(J_0)\, (i+1)^{-\beta} \leq \frac{\delta_2 \sqrt{\epsilon}}{2\,b_\nabla(J_0)}$;
    \item $\displaystyle\sum_{i=0}^\infty \eta_i^2 \leq \frac{r(J_0)^2 \delta_3}{c(J_0)}$, i.e., condition \eqref{eq:lemC-condition} of Lemma~\ref{lem:event-C} with $\delta = \delta_3$.
\end{enumerate}
Then the event $F =\Omega  \cap C$ defined in \eqref{eventeqf} occurs with probability at least $1-\delta$.
\end{Lemma}
By Lemma~\ref{lem:event-F}, with probability at least $1-\delta$, the iterates remain in $\mathcal{S}(J_0)$ and consecutive iterates are within distance $r(J_0)$. For polynomially decaying step sizes, its conditions take the following form.
\begin{Corollary}\label{cor:event-F}
Suppose that Definition~\ref{def:local-smooth} and Assumptions~\ref{as:grad-bound} and~\ref{as:oracle} hold, and let $\epsilon$, $\delta_1,\delta_2,\delta_3$, $\delta$ and $\sigma_\zeta^2(J_0)$ be as in Lemma~\ref{lem:event-F}. Let the step sizes satisfy $\underline{\eta}\,(i+1)^{-\kappa}\leq\eta_i\leq\bar{\eta}\,(i+1)^{-\kappa}$ for all $i\in\mathbb{Z}_+$, with $\kappa\in(\frac{1}{2},1)$, $\kappa+\beta>1$ and constants $0<\underline{\eta}\leq\bar{\eta}$, and
\begin{subequations}\label{stepsize}
\begin{align}
    &\sum_{i=0}^{\infty}\eta_i^2\leq\min\left\{\frac{\delta_1\epsilon}{\sigma_\zeta^2(J_0)+\frac{1}{2}L(J_0)\,c(J_0)},\ \frac{r(J_0)^2\,\delta_3}{c(J_0)}\right\},\label{stepsize-b}\\
    &\sum_{i=0}^{\infty}\eta_i\,\Delta_0(J_0)\,(i+1)^{-\beta}\leq\frac{\delta_2\sqrt{\epsilon}}{2\,b_\nabla(J_0)}.\label{biasterm}
\end{align}
\end{subequations}
Then $\mathbb{P}(F)\geq1-\delta$. 
\end{Corollary}

Corollary~\ref{cor:event-F} follows from Lemma~\ref{lem:event-F}, since \eqref{stepsize-b}--\eqref{biasterm} are its conditions~1--3; both series converge since $2\kappa>1$ and $\kappa+\beta>1$, so that \eqref{stepsize} holds for all sufficiently small $\bar\eta$. On $F$, the descent inequality of Definition~\ref{def:local-smooth} holds with the uniform constant $L(J_0)$, and combining it with gradient domination yields a recursion for the optimality gap.

\begin{Theorem}\label{thm:main-convergence}
Suppose that Definition~\ref{def:GD} holds and that the conditions of Corollary~\ref{cor:event-F} are satisfied with $\bar\eta<\mu(J_0)$, so that $\mathbb{P}(F)\geq1-\delta$. Then the following hold:
\begin{enumerate}
\item For every $i\in\mathbb{Z}_+$,
\begin{equation}\label{finite}
\begin{split}
    \mathbb{E}\big[(f(X_i)-f(X^*))\boldsymbol{1}_F\big]\leq{}&(f(X_0)-f(X^*))\,e^{-\frac{\underline{\eta}}{\mu(J_0)(1-\kappa)}\left((i+1)^{1-\kappa}-1\right)}\\+\gamma_1\,e^{-\gamma(i+1)^{1-\kappa}}
    &+\gamma_2\,(i+1)^{-\beta}+\gamma_3\,(i+1)^{-\kappa},
\end{split}
\end{equation}
with $\nu(J_0):=c(J_0)\big(\frac{L(J_0)}{2}+\frac{b_\nabla(J_0)}{r(J_0)}\big)$; $\gamma:=\frac{\underline{\eta}\,(1-(3/4)^{1-\kappa})}{\mu(J_0)(1-\kappa)}$; $\gamma_3:=2^{\kappa}\mu(J_0)\,\bar{\eta}\,\nu(J_0)$;\\$\gamma_2:=2^{\beta}\mu(J_0)\,b_\nabla(J_0)\Delta_0(J_0)$; $\gamma_1:=\bar{\eta}\,b_\nabla(J_0)\Delta_0(J_0)\big(1+\frac{1}{\kappa+\beta-1}\big)+\bar{\eta}^2\nu(J_0)\big(1+\frac{1}{2\kappa-1}\big)$.
\item For any $\lambda\in(\max\{2-2\kappa,2-\kappa-\beta\},1)$,
\begin{equation}\label{aresult}
    (f(X_i)-f(X^*))\boldsymbol{1}_F=o\big(i^{-(1-\lambda)}\big)\quad\text{almost surely}.
\end{equation}
\end{enumerate}
\end{Theorem}
\begin{proof}[Proof of Theorem~\ref{thm:main-convergence}]
By Corollary~\ref{cor:event-F}, we have $\mathbb{P}(F)\geq1-\delta$. Recall that $\hat g_i:=\hat\nabla f(X_i,\xi_i)$, so that $X_{i+1}=X_i-\eta_i\hat g_i$, and that the events $F_i$ in \eqref{eq:event-Fi} are $\mathcal{F}_i$-measurable with $F_{i+1}\subseteq F_i$ and $\boldsymbol{1}_F\leq\boldsymbol{1}_{F_i}$. Let $D_i:=f(X_i)-f(X^*)$, $B_i:=\{\|X_{i+1}-X_i\|\leq r(J_0)\}$, $Y_i:=D_i\boldsymbol{1}_{F_i}$ and $a_i:=\mathbb{E}[Y_i]$, and write $b:=b_\nabla(J_0)$, $L:=L(J_0)$, $r:=r(J_0)$, $\mu:=\mu(J_0)$, $c:=c(J_0)$ and $\nu:=\nu(J_0)$.

\emph{Step 0 (one-step recursion).} On $F_i\cap B_i$, we have $f(X_i)\leq J_0$ and $\|X_{i+1}-X_i\|\leq r\leq r(f(X_i))$, so Definition~\ref{def:local-smooth} gives
\begin{equation}\label{eq:descent}
    D_{i+1}\leq D_i-\eta_iT_i+\frac{L}{2}\eta_i^2\|\hat g_i\|^2,\qquad T_i:=\langle \hat g_i, \nabla f(X_i) \rangle.
\end{equation}
Since $F_{i+1}\subseteq F_i\cap B_i$, $D_{i+1}\geq0$ and $D_i\boldsymbol{1}_{F_i\cap B_i}\leq Y_i$, it follows that $Y_{i+1}\leq Y_i-\eta_iT_i\boldsymbol{1}_{F_i}+\eta_i|T_i|\boldsymbol{1}_{F_i}\boldsymbol{1}_{B_i^c}+\frac{L}{2}\eta_i^2\|\hat g_i\|^2\boldsymbol{1}_{F_i}$. On $F_i$, Definition~\ref{def:GD} and Assumptions~\ref{as:grad-bound} and~\ref{as:oracle} give
\begin{align*}
    \mathbb{E}[T_i\mid\mathcal{F}_i]&=\|\nabla f(X_i)\|^2+\langle \Delta_i, \nabla f(X_i) \rangle\geq\frac{D_i}{\mu}-b\,\|\Delta_i\|,\\
    \mathbb{E}\big[|T_i|\boldsymbol{1}_{B_i^c}\mid\mathcal{F}_i\big]&\leq b\,\mathbb{E}\big[\|\hat g_i\|^2\mid\mathcal{F}_i\big]^{1/2}\,\mathbb{P}(B_i^c\mid\mathcal{F}_i)^{1/2}\leq\frac{b\,c\,\eta_i}{r},\qquad\mathbb{E}\big[\|\hat g_i\|^2\mid\mathcal{F}_i\big]\leq c,
\end{align*}
where the second line uses the Cauchy--Schwarz inequality and $\mathbb{P}(B_i^c\mid\mathcal{F}_i)=\mathbb{P}(\|\hat g_i\|\\>r/\eta_i\mid\mathcal{F}_i)\leq c\,\eta_i^2/r^2$ by Markov's inequality. Hence
\begin{equation}\label{eq:one-step-cond}
    \mathbb{E}[Y_{i+1}\mid\mathcal{F}_i]\leq\Big(1-\frac{\eta_i}{\mu}\Big)Y_i+z_i,\qquad z_i:=\eta_i\big(b\,\|\Delta_i\|\boldsymbol{1}_{F_i}+\eta_i\,\nu\big),
\end{equation}
where $z_i\geq0$ is $\mathcal{F}_i$-measurable. Since $F_i\subseteq\{X_i\in\mathcal{S}(J_0)\}$, \eqref{eq:bias-decay} gives
\begin{equation}\label{eq:zbar}
    \mathbb{E}[z_i]\leq\bar z_i:=\eta_i\big(b\,\Delta_0(J_0)(i+1)^{-\beta}+\eta_i\,\nu\big),
\end{equation}
and, taking full expectations in \eqref{eq:one-step-cond},
\begin{equation}\label{eq:one-step-exp}
    a_{i+1}\leq\Big(1-\frac{\eta_i}{\mu}\Big)a_i+\bar z_i,\qquad\forall i\in\mathbb{Z}_+.
\end{equation}

\emph{Part I (rate in expectation).} Define $\Pi_{i,k}:=\prod_{j=k}^{i-1}(1-\eta_j/\mu)$, with $\Pi_{i,i}:=1$. Iterating \eqref{eq:one-step-exp} gives
\begin{equation}\label{eq:unrolled}
    a_i\leq\Pi_{i,0}\,a_0+\sum_{k=0}^{i-1}\Pi_{i,k+1}\bar z_k,\qquad\forall i\in\mathbb{Z}_+.
\end{equation}
Since $0\leq1-\eta_j/\mu\leq e^{-\eta_j/\mu}$ by $\eta_j\leq\bar\eta<\mu$, $\eta_j\geq\underline{\eta}(j+1)^{-\kappa}$, and $\sum_{j=k}^{i-1}(j+1)^{-\kappa}\geq\int_{k+1}^{i+1}x^{-\kappa}\,dx$, we have
\begin{equation}\label{eq:Pi-bound}
    \Pi_{i,k}\leq\exp\Big(-\frac{\underline{\eta}}{\mu(1-\kappa)}\big[(i+1)^{1-\kappa}-(k+1)^{1-\kappa}\big]\Big).
\end{equation}
With $k=0$ and $a_0=f(X_0)-f(X^*)$, \eqref{eq:Pi-bound} bounds the first term in \eqref{eq:unrolled} by the first term in \eqref{finite}. We split the sum in \eqref{eq:unrolled} at $M:=\lfloor i/2\rfloor$, and note that $M+1\geq(i+1)/2$, and $M+1\leq\frac{3}{4}(i+1)$ whenever $M\geq1$.

\emph{Early segment ($k<M$).} Since $\Pi_{i,k+1}\leq\Pi_{i,M}$, \eqref{eq:Pi-bound} and $M+1\leq\frac{3}{4}(i+1)$ give $\Pi_{i,k+1}\leq e^{-\gamma(i+1)^{1-\kappa}}$. Moreover, since $\kappa+\beta>1$ and $2\kappa>1$, $\sum_{k=0}^{M-1}(k+1)^{-(\kappa+\beta)}\leq1+\frac{1}{\kappa+\beta-1}$ and $\sum_{k=0}^{M-1}(k+1)^{-2\kappa}\leq1+\frac{1}{2\kappa-1}$. With $\eta_k\leq\bar{\eta}(k+1)^{-\kappa}$, this gives $\sum_{k=0}^{M-1}\bar z_k\leq\gamma_1$, and hence $\sum_{k=0}^{M-1}\Pi_{i,k+1}\bar z_k\leq\gamma_1\,e^{-\gamma(i+1)^{1-\kappa}}$.

\emph{Late segment ($k\geq M$).} Since $\Pi_{i,k+1}-\Pi_{i,k}=\frac{\eta_k}{\mu}\Pi_{i,k+1}$, 
\begin{equation}\label{eq:telescope}
    \sum_{k=M}^{i-1}\Pi_{i,k+1}\,\eta_k=\mu\big(1-\Pi_{i,M}\big)\leq\mu.
\end{equation}
For $k\geq M$, we have $(k+1)^{-\beta}\leq(M+1)^{-\beta}\leq2^{\beta}(i+1)^{-\beta}$ and $\eta_k\leq\bar{\eta}(M+1)^{-\kappa}\leq2^{\kappa}\bar{\eta}(i+1)^{-\kappa}$. Together with \eqref{eq:telescope}, this gives
\begin{equation*}
\begin{split}
       \sum_{k=M}^{i-1}\Pi_{i,k+1}\bar z_k&\leq\Big(2^{\beta}b\,\Delta_0(J_0)(i+1)^{-\beta}+2^{\kappa}\bar{\eta}\,\nu\,(i+1)^{-\kappa}\Big)\sum_{k=M}^{i-1}\Pi_{i,k+1}\eta_k\\&\leq\gamma_2(i+1)^{-\beta}+\gamma_3(i+1)^{-\kappa}.
\end{split}
\end{equation*}
Substituting the three bounds into \eqref{eq:unrolled} and using $\mathbb{E}[D_i\boldsymbol{1}_F]\leq a_i$ yields item~1.

\emph{Part II (almost-sure rate).} Let $s:=1-\lambda\in(0,\min\{2\kappa-1,\kappa+\beta-1\})$ and $V_i:=(i+1)^sY_i$, which is non-negative and $\mathcal{F}_i$-measurable. By \eqref{eq:one-step-cond} and $((i+2)/(i+1))^s\leq1+s/(i+1)$,
\begin{equation*}
    \mathbb{E}[V_{i+1}\mid\mathcal{F}_i]\leq\Big(1+\frac{s}{i+1}\Big)\Big(1-\frac{\eta_i}{\mu}\Big)V_i+(i+2)^sz_i\leq\Big(1-\frac{\eta_i}{\mu}+\frac{s}{i+1}\Big)V_i+(i+2)^sz_i.
\end{equation*}
Since $\kappa<1$, there exists $i_1\in\mathbb{Z}_+$ such that $s(i+1)^{\kappa-1}\leq\underline{\eta}/(2\mu)$, and hence $s/(i+1)\leq\eta_i/(2\mu)$, for all $i\geq i_1$. Therefore, $\mathbb{E}[V_{i+1}\mid\mathcal{F}_i]\leq V_i-\frac{\eta_i}{2\mu}V_i+(i+2)^sz_i$ for all $i\geq i_1$.

By \eqref{eq:zbar}, $\mathbb{E}[(i+2)^sz_i]\leq(i+2)^s\bar z_i\leq2^s\big(\bar{\eta}\,b\,\Delta_0(J_0)(i+1)^{s-(\kappa+\beta)}+\bar{\eta}^2\nu\,(i+1)^{s-2\kappa}\big)$. Since $s<\min\{\kappa+\beta-1,2\kappa-1\}$, this gives $\mathbb{E}\big[\sum_{i}(i+2)^sz_i\big]<\infty$, and hence $\sum_{i}(i+2)^sz_i<\infty$ almost surely. The Robbins--Siegmund theorem~\cite{ROBBINS1971233}, applied from $i_1$ on with the non-negative $\mathcal{F}_i$-measurable terms $(i+2)^sz_i$, then implies that $V_i$ converges almost surely to a finite limit $V_\infty$ and that $\sum_{i}\eta_iV_i<\infty$ almost surely. Since $\sum_i\eta_i=\infty$, we have $\liminf_{i}V_i=0$, hence $V_\infty=0$ almost surely. Item~2 follows from $(i+1)^s(f(X_i)-f(X^*))\boldsymbol{1}_F\leq V_i$.
\end{proof}

The following observations are in order:
\begin{enumerate}
    \item Theorem~\ref{thm:main-convergence} establishes that, on the event $F$, which has probability at least $1-\delta$, the optimality gap $f(X_i)-f(X^*)$ vanishes both in expectation and almost surely. Item~1 bounds the expected optimality gap at every iteration with explicit constants: the first two terms in \eqref{finite} form a transient that decays as $e^{-\gamma(i+1)^{1-\kappa}}$, while $\gamma_2(i+1)^{-\beta}$ and $\gamma_3(i+1)^{-\kappa}$ are the contributions of the oracle bias and of the gradient noise, respectively. Hence $\mathbb{E}[(f(X_i)-f(X^*))\boldsymbol{1}_F]=O\big(i^{-\min\{\kappa,\beta\}}\big)$. Item~2 gives an almost-sure rate, which does not follow from item~1.
    \item The rate in expectation is limited by the gradient noise if $\kappa<\beta$ and by the bias if $\kappa>\beta$. Since $\gamma_2$ does not depend on the step size, a larger $\kappa$ does not improve the rate in the bias-limited regime and only slows the transient. The almost-sure rate is slower, since it is limited by the summability of the forcing terms in the Robbins--Siegmund argument.
    \item The condition $\kappa>\frac{1}{2}$ ensures $\sum_i\eta_i^2<\infty$, and $\kappa+\beta>1$ ensures that the accumulated bias $\sum_i\eta_i\Delta_0(J_0)(i+1)^{-\beta}$ is finite. Since $\kappa\leq1$, $\sum_i\eta_i=\infty$, which is needed for the optimality gap to vanish; the strict inequality $\kappa<1$ yields the transient $e^{-\gamma(i+1)^{1-\kappa}}$ and is used in the proof of item~2. In \eqref{stepsize}, $\epsilon$ is the margin that keeps the iterates in $\mathcal{S}(J_0)$, and $\delta_1$, $\delta_2$ and $\delta_3$ split the probability $\delta$ that $F$ fails among the gradient noise, the bias, and steps longer than $r(J_0)$, respectively.
\end{enumerate}
\section{SGD-based Analysis of Policy Gradient for the LQR}\label{sec:LQR-example}
Having developed, in Sections~\ref{sec:abstract-setting}--\ref{sec:Convergence}, a convergence theory for SGD on a general objective $f:\mathcal X\to\mathbb R\cup\{+\infty\}$ under local smoothness, local gradient domination, and a biased oracle, we now instantiate it on the averaged infinite-horizon LQR problem. To this end, we take $\mathcal X:=\mathbb R^{n_u\times n_x}$ with $\langle K,K'\rangle:=\operatorname{Tr}(K^\top K')$, and identify $X$, $f$ and $\mathcal S(J)$ with the feedback gain $K$, the LQR cost $C$, extended by $C(K):=+\infty$ for $K\notin\mathcal S$, and its sublevel set $\mathcal S(J)$. We derive conditions under which Assumption~\ref{as:oracle} holds, with explicit constants, for two gradient estimators, and thereby obtain the guarantees of Theorem~\ref{thm:main-convergence}.
\subsection{LQR Problem Setup}\label{sec:PSP}
We consider the following averaged infinite-horizon optimal control problem, where the plant is subject to additive stochastic noise:
\begin{subequations}\label{LTI}
\begin{align}
        \min_{\{u_t\}}& \lim_{T\rightarrow +\infty}\frac{1}{T}\mathop{\mathbb{E}}_{x_0, w_t} \sum_{t=0}^{T-1}\left( x_t^\top Q x_t +u_t ^\top R u_t\right),\\
        \mathrm{s.t.} ~ x_{t+1}=&Ax_t+Bu_t+w_t, \label{system}x_0 \sim \mathcal{N}(0,\Sigma_0), w_t\sim \mathcal{N}(0,\Sigma_w),
\end{align}
\end{subequations}
where $A\in \mathbb{R}^{n_x \times n_x}$, $B\in \mathbb{R}^{n_x \times n_u}$, $(A,B)$ is stabilizable but unknown; covariance matrices $\Sigma_0,\Sigma_w\succ0$; $Q,R\succ 0$ are the weight matrices. We define the set of stabilizing feedback gains as $\mathcal{S} := \left\{ K \in \mathbb{R}^{n_u \times n_x} \,\big|\, \rho(A_K)< 1 \right\},~\text{with}~ A_K:= A + B K.$ The infinite-horizon average cost under a linear policy $u_t = K x_t$ with $K \in \mathcal{S}$ is given by
\begin{equation}\label{CostFunction}
    C(K) := \lim_{T \to \infty} \frac{1}{T} \mathbb{E}_{x_0, w_t} \left[ \sum_{t=0}^{T-1} x_t^\top Q_K x_t \right],\qquad \text{with}~ Q_K := Q + K^\top R K.
\end{equation}
For any $K \in \mathcal{S}$, the gradient of $C$ is given by
\begin{equation}\label{Gradient}
    \nabla C(K) = 2 E_K \Sigma_K,
\end{equation}
where $E_K := \left( R + B^\top P_K B \right) K + B^\top P_K A$, $P_K$ solves the Lyapunov equation $P_K=A_K^\top P_K A_K+Q_K$, and $\Sigma_K$ is the average state covariance $\Sigma_K:=\lim_{T\rightarrow +\infty}\frac{1}{T}\sum_{t=0}^{T-1} \Sigma_t,$ $\Sigma_t:=\mathop{\mathbb{E}}_{x_0,w_t}[x_t x_t^\top].$ It is well known \cite{lewis2012optimal} that the optimal $K^*$ minimizing $C$ satisfies
\begin{subequations}
    \begin{align}
K^*&=-(R+B^\top P_{K^*}B)^{-1}B^\top P_{K^*}A, \label{Kpolicyimprovement} \\
P_{K^*}&=Q+A^\top P_{K^*}A -A^\top P_{K^*}B(R+B^\top P_{K^*}B)^{-1}B^\top P_{K^*}A. \label{Ppolicyimprovement}
\end{align}
\end{subequations}
Consistent with the sublevel set of Definition~\ref{def:level-set}, for $J\geq C(K^*)$, we define
\begin{equation}\label{levelset}
    \mathcal{S}(J) := \left\{K \in \mathbb{R}^{n_u\times n_x} ~\big|~ C(K) \leq J \right\}.
\end{equation}
We now recall the properties of $C$, especially gradient domination and local smoothness.
\begin{Lemma}[Boundedness of $\lVert\nabla C(K)\rVert, \lVert K\rVert$]\cite[Lemma 22]{pmlr-v80-fazel18a} \label{bound}
Given any $J_0\geq C(K^*)$, for all $K\in \mathcal{S}(J_0)$, we have
\begin{equation}
        \lVert \nabla C(K) \rVert_F\leq b_{\nabla}(J_0), \quad \lVert K \rVert \leq b_K(J_0),
\end{equation}
where $b_\nabla,b_K$ are given in \eqref{boundedgradienteq} and \eqref{boundK} in Appendix~\ref{DetailedExpression}, respectively.
\end{Lemma}
Lemma~\ref{bound} shows that $C$ satisfies Assumption~\ref{as:grad-bound} with the explicit, non-decreasing bound $b_\nabla(\cdot)$ given in \eqref{boundedgradienteq}.
\begin{Lemma}[Local Lipschitz Continuity of $\Sigma_K,C,\nabla C$]\cite[Lemmas 16, 24, 25]{pmlr-v80-fazel18a}\label{Lipschitz}
    Suppose $K\in\mathcal S$ and $\|K-K'\|\le h(C(K))$, with 
    \begin{equation}
     h(C(K)):=\frac{\lambda_1(\Sigma_w)\lambda_1(Q)}{4C(K)\lVert B \rVert(\lVert A\rVert+\lVert B \rVert b_K(C(K))+1)}.
    \end{equation}
    Then $K'\in\mathcal S$ and $\lVert \Sigma_K-\Sigma_{K'} \rVert \leq h_{\Sigma}(C(K))\lVert K-K' \rVert$ with $h_{\Sigma}(C(K)):=\frac{C(K)}{\lambda_1(Q)h(C(K))}$. If, in addition, $\lVert K-K' \rVert\leq \min\{h(C(K)),\tilde{c}\}$ for some fixed $\tilde c>0$, then
    \begin{subequations}
    \begin{align}
        \lvert C(K)-C(K')\rvert \leq h_{C}(C(K))\lVert K-K' \rVert,\\
        \lVert \nabla C(K)-\nabla C(K') \rVert_F \leq h_{\nabla}(C(K)) \lVert K-K' \rVert,
    \end{align}
\end{subequations}
where $h_{C},h_{\nabla}$ are defined in \eqref{Pertubation1} and \eqref{ErrorGradient} in Appendix~\ref{DetailedExpression}, respectively.
\end{Lemma}
\begin{Lemma}[Gradient Domination]\cite[Lemma 3]{pmlr-v80-fazel18a}\label{LemmaGD}
    For any $K \in \mathcal{S}$,
    \begin{equation}\label{GDMu}
        C(K)-C(K^*)\leq \mu\lVert \nabla C(K)\rVert_F^2,\quad\mathrm{with~}\mu:=\tfrac{1}{4}\lVert \Sigma_{K^*} \rVert\lVert \Sigma_w ^{-2}\rVert\lVert R ^{-1}\rVert.
    \end{equation}
\end{Lemma}
Lemma~\ref{LemmaGD} shows that $C$ satisfies Definition~\ref{def:GD} with $\mu(J)\equiv\mu$. The property is local, since it only holds on the set $\mathcal{S}$ of stabilizing gains, outside of which $C$ is infinite.
\begin{Lemma}[Local Smoothness]\cite[Lemmas 12, 16]{pmlr-v80-fazel18a}\label{quasi}
    For any $K\in\mathcal S$ and $K'$ with $\|K'-K\|_F\le r(C(K))$, we have $K'\in\mathcal S$ and
    \begin{equation}\label{Smoothness1}
        \left| C(K') - C(K) - \mathrm{Tr}\left( (K' - K)^\top \nabla C(K) \right) \right|
        \leq \frac{L(C(K))}{2} \| K' - K \|_F^2,
    \end{equation}
    where
    \begin{subequations}
        \begin{align}
            L(C(K))&:=\frac{64C(K)}{\lambda_1(Q)\lambda_1(\Sigma_w)}\left( \lVert B \rVert^2 C(K)+\lambda_1(\Sigma_w)\lVert R \rVert \right),\\
            r(C(K))&:=\frac{\lambda_1(Q)^2 \lambda_1(\Sigma_w)^2}{32\lVert B \rVert C(K)^2(1+\lVert A\rVert+\lVert B\rVert b_K(C(K)))}.\label{expressionr}
        \end{align}
    \end{subequations}
\end{Lemma}
 Lemma~\ref{quasi} is precisely the LQR instance of Definition~\ref{def:local-smooth}. Together, Lemmas~\ref{bound}, \ref{LemmaGD}, and \ref{quasi} confirm that the LQR cost $C$ is a valid instance of the abstract objective $f$ satisfying gradient domination, local smoothness, and bounded gradient, with all constants ($b_\nabla,\mu,L,r$) given explicitly in terms of the problem data $(A,B,Q,R,\Sigma_w)$.

\subsection{Stochastic Gradient Oracles from Indirect and Direct Data-driven Algorithms}\label{Sec:GradientOracle}

We study two data-driven estimators of $\nabla C(K)$ when $(A,B)$ is unknown: an indirect one, which identifies the system matrices by recursive least squares (Section~\ref{Sec:Indirect}), and a direct one, which estimates the gradient from empirical cost evaluations by a zeroth-order method (Section~\ref{sec:direct}). Both are built from trajectory data of \eqref{system} and play the role of the oracle $\hat\nabla f(X,\xi)$ in \eqref{eq:sgd-iteration}; for each, we characterize its bias and second moment as in \eqref{eq:oracle}.

\subsubsection{Indirect Gradient Oracle}\label{Sec:Indirect}
The indirect oracle evaluates the gradient formula \eqref{Gradient} at the RLS estimate of $(A,B)$. We characterize this oracle here and analyze the resulting closed-loop scheme in Section~\ref{sec:indirectclosed}.

Let $\theta:=[A,B]$ and let $d_t:=[x_t^\top,u_t^\top]^\top$ denote the regressor, so that $x_{t+1}=\theta d_t+w_t$. Throughout, the inputs contain a dithering signal $e_t\sim\mathcal{N}(0,\Sigma_e)$, $\Sigma_e\succ0$, i.i.d.\ and independent of $\{w_t\}$ and of the initial state. Given $n$ samples $\{(d_t,x_{t+1})\}_{t=0}^{n-1}$, the least-squares estimate $\hat\theta=[\hat A,\hat B]$ and its error $\Delta\theta:=\hat\theta-\theta$ are
\begin{equation}\label{eq:LS}
    \hat\theta=\sum_{t=0}^{n-1}x_{t+1}d_t^\top H^{-1},\qquad
\Delta\theta=\sum_{t=0}^{n-1}w_td_t^\top H^{-1},\qquad
H:=\sum_{t=0}^{n-1}d_td_t^\top,
\end{equation}
whenever $H\succ0$. When a new sample $(d,x^+)$ is collected, the estimate is updated recursively by
\begin{equation}\label{eq:RLS}
    H^+=H+dd^\top,\qquad
    \hat\theta^+=\hat\theta+(x^+-\hat\theta d)\,d^\top(H^+)^{-1}.
\end{equation}
The gradient constructed from the estimated model is
\begin{equation}\label{eq:indirect-grad}
    \hat\nabla_IC(K,\hat A,\hat B):=2\hat E_K\hat\Sigma_K,\qquad
    \hat E_K:=\big(R+\hat B^\top\hat P_K\hat B\big)K+\hat B^\top\hat P_K\hat A,
\end{equation}
where $\hat P_K$ and $\hat\Sigma_K$ solve $\hat P_K=(\hat A+\hat BK)^\top\hat P_K(\hat A+\hat BK)+Q_K$ and $\hat\Sigma_K=(\hat A+\hat BK)\hat\Sigma_K(\hat A+\hat BK)^\top+\Sigma_w$; it is well defined whenever $\hat A+\hat BK$ is Schur stable.

The resulting scheme is summarized in Algorithm~\ref{alg:indirect}. After an offline phase of $t_0$ samples generated with the initial gain $K_0$, it alternates between a gradient step at the current gain $K_i$, computed with the estimate $\hat\theta_i$, and the collection of one new sample, which updates the estimate via \eqref{eq:RLS}. The data are generated with an excitation gain that either remains at the initial gain, $K^{\rm ex}_i=K_0$ (off-policy), or coincides with the current iterate, $K^{\rm ex}_i=K_i$ (on-policy); in particular, the gain at which the gradient is evaluated need not coincide with the gains that generate the data. In the remainder of this subsection, we bound the gradient error in terms of the estimation error $\Delta\theta$, and quantify the decay of $\Delta\theta$ as data accumulate.

\begin{algorithm}[t]
\caption{Indirect data-driven policy gradient}\label{alg:indirect}
\begin{algorithmic}
\REQUIRE $K_0\in\mathcal{S}$; step sizes $\{\eta_i\}$; offline length $t_0$; $\Sigma_e\succ0$.
\FOR{$t=0,\dots,t_0-1$}
\STATE Apply $u_t=K_0x_t+e_t$ and collect $d_t$ and $x_{t+1}$.
\ENDFOR
\STATE Compute $\hat\theta_0=[\hat A_0,\hat B_0]$ and $H_0$ via \eqref{eq:LS} from the $t_0$ samples.
\FOR{$i=0,1,\dots$}
\STATE Compute $\hat g_i:=\hat\nabla_IC(K_i,\hat A_i,\hat B_i)$ via \eqref{eq:indirect-grad}.
\STATE Update $K_{i+1}=K_i-\eta_i\hat g_i$.
\STATE Set $K^{\rm ex}_{i+1}:=K_0$ (off-policy) or $K^{\rm ex}_{i+1}:=K_{i+1}$ (on-policy).
\STATE Apply $u_{t_0+i}=K^{\rm ex}_{i+1}x_{t_0+i}+e_{t_0+i}$, collect $d_{t_0+i}$ and $x_{t_0+i+1}$, and compute $(\hat\theta_{i+1},H_{i+1})$ via \eqref{eq:RLS}.
\ENDFOR
\end{algorithmic}
\end{algorithm}

The following lemma bounds the gradient error induced by $\Delta\theta$, uniformly over a sublevel set.

\begin{Lemma}[Gradient Error]\label{lem:grad-error}
Let $J_0\geq C(K^*)$, and define
\begin{equation}\label{eq:ptheta-uniform}
    p_\theta:=\frac{\lambda_1(Q)\,\lambda_1(\Sigma_w)}{4J_0\big(1+\|A\|+\|B\|b_K(J_0)\big)\big(1+b_K(J_0)\big)},\qquad
    \bar g(J_0):=b_\nabla(J_0)+p\,p_\theta,
\end{equation}
with $p$ defined in \eqref{eq:p-def}. Then, for every $K\in\mathcal{S}(J_0)$ and every estimate with $\|\Delta\theta\|_F\leq p_\theta$, the matrix $\hat A+\hat BK$ is Schur stable and
\begin{equation}\label{eq:grad-error}
    \big\|\hat\nabla_IC(K,\hat A,\hat B)-\nabla C(K)\big\|_F\leq p\,\|\Delta\theta\|_F,\qquad
    \big\|\hat\nabla_IC(K,\hat A,\hat B)\big\|_F\leq\bar g(J_0).
\end{equation}
\end{Lemma}

Lemma~\ref{lem:grad-error} directly yields the following characterization of the oracle.

\begin{Lemma}[Gradient Oracle from Indirect Method]\label{lem:indirect-oracle}
Let $K\in\mathcal S(J_0)$, let $\hat\theta$ satisfy $\|\Delta\theta\|_F\le p_\theta$, and suppose that $K$ and $\hat\theta$ are measurable with respect to a $\sigma$-algebra $\mathcal F$, e.g., the one generated by the data used to compute $\hat\theta$. Then $\hat\nabla_I C(K,\hat A,\hat B)$ is well defined, $\mathcal F$-measurable, and satisfies
\begin{subequations}\label{eq:indirect-oracle}
\begin{align}
    \mathbb{E}\big[\hat\nabla_IC(K,\hat A,\hat B)\mid\mathcal{F}\big]&=\nabla C(K)+\Delta_I,\qquad\|\Delta_I\|_F\leq p\,\|\Delta\theta\|_F,\label{eq:indirect-oracle-mean}\\
    \mathbb{E}\big[\|\hat\nabla_IC(K,\hat A,\hat B)\|_F^2\mid\mathcal{F}\big]&\leq\bar g(J_0)^2,\label{eq:indirect-oracle-var}
\end{align}
\end{subequations}
where $\Delta_I:=\hat\nabla_IC(K,\hat A,\hat B)-\nabla C(K)$.
\end{Lemma}

Both claims follow directly from Lemma~\ref{lem:grad-error}. Since the indirect oracle is computed from past data only, its whole error is a bias $\Delta_I$, which varies with the data through $\Delta\theta$. 
Its accuracy hinges on the decay of $\Delta\theta$, which depends on the second moment of the data and on their informativity. The next lemma bounds the second moment when the gains remain in a sublevel set and vary slowly.

\begin{Lemma}[Mean-square Boundedness under Slowly Varying Gains]\label{lem:ms-bound}
Let $J_0\geq C(K^*)$, and consider the closed loop $x_{t+1}=(A+B\bar K_t)x_t+Be_t+w_t$, $t\in\mathbb{Z}_+$, with $x_0\sim\mathcal{N}(0,\Sigma_0)$, where $\{\bar K_t\}$ is a sequence of excitation gains. Define $\chi(J_0):=\sqrt{\frac{J_0}{\lambda_1(Q)\lambda_1(\Sigma_w)}}$, $\varrho(J_0):=1-\sqrt{1-\chi(J_0)^{-2}}$, and $\varepsilon_K(J_0):=\min\Big\{h(J_0),\ \frac{\varrho(J_0)\,\lambda_1(\Sigma_w)}{h_\Sigma(J_0)}\Big\}$ with $h$, $h_\Sigma$ as in Lemma~\ref{Lipschitz}. If, surely, $\bar K_t\in\mathcal{S}(J_0)$ and $\|\bar K_{t+1}-\bar K_t\|\leq\varepsilon_K(J_0)$ for all $t\in\mathbb{Z}_+$, then
\begin{equation}\label{eq:ms-bound}
    \sup_{t\in\mathbb{Z}_+}\mathbb{E}\big[\|x_t\|^2\big]\leq\bar x(J_0):=\chi(J_0)^2\Big(\sqrt{\mathrm{Tr}(\Sigma_0)}+\frac{2}{\varrho(J_0)}\sqrt{\mathrm{Tr}(\Sigma_w)+\mathrm{Tr}(B\Sigma_eB^\top)}\Big)^2.
\end{equation}
\end{Lemma}

Informativity of the data is formalized as follows.
\begin{Assumption}[Persistency of Excitation]\label{as:PE}
There exist $\alpha>0$, $\delta_5\in(0,1)$, and $T_{\rm PE}\in\mathbb{Z}_{++}$ such that, whenever each excitation gain $\bar K_t$ is independent of $\{(w_s,e_s)\}_{s\ge t}$ and $\{\bar K_t\}$ satisfies the conditions of Lemma~\ref{lem:ms-bound},
\begin{equation*}
    \mathbb{P}\Big(\textstyle\sum_{t=0}^{s-1}d_td_t^\top\succeq\alpha s\,I_{n_x+n_u}\ \text{ for all } s\ge T_{\rm PE}\Big)\ge1-\delta_5.
\end{equation*}
\end{Assumption}
For $t_0\ge T_{\rm PE}$ and $H_i=\sum_{t=0}^{t_0+i-1}d_td_t^\top$ as in Algorithm~\ref{alg:indirect}, we define the excitation events
\begin{equation}\label{eq:PE-events}
    \mathrm{PE}_i:=\big\{H_k\succeq\alpha(t_0+k)I_{n_x+n_u},\ k=0,\dots,i\big\},\qquad \mathrm{PE}:=\textstyle\bigcap_{i\in\mathbb{Z}_+}\mathrm{PE}_i .
\end{equation}
$\mathrm{PE}_i$ is determined by the data up to time $t_0+i-1$, and $\mathbb{P}(\mathrm{PE})\ge1-\delta_5$ under the conditions of Assumption~\ref{as:PE}. The threshold $T_{\rm PE}$ makes this probability independent of $t_0$. Assumption~\ref{as:PE} is implied by a high-probability version of the local persistency condition of~\cite[Definition~2]{https://doi.org/10.1002/rnc.7475}. 

Assumption~\ref{as:PE} is not analyzed here, but it is plausible: at every time step, the fresh noise and dithering excite the state and the input independently of the gains, so that, conditionally on the past, the new state and dithering sample are Gaussian with a nondegenerate covariance. In the off-policy case, where the data are generated by a fixed gain, the assumption can be established by the small-ball argument of~\cite[Proposition~2.5 and Lemma~4.1]{simchowitz2018learning}, combined with a union bound over dyadic times; the resulting excitation level scales with the smallest eigenvalue of the noise and dithering covariances, and the required threshold grows only logarithmically in the inverse failure probability. The on-policy case, in which the data depend on the gains and hence on past noise, requires additional arguments and is left to future work.

The next theorem bounds the estimation error on the excitation events \eqref{eq:PE-events}.

\begin{Theorem}[Estimation Error]\label{thm:RLS}
Suppose that each excitation gain $\bar K_t$ is independent of $\{(w_s,e_s)\}_{s\ge t}$ and that $\{\bar K_t\}$ satisfies the conditions of Lemma~\ref{lem:ms-bound}. Let $t_0\ge T_{\rm PE}$, and let $\hat\theta_i$ be the estimate \eqref{eq:LS} from the first $t_0+i$ samples. Then, for all $i\in\mathbb{Z}_+$,
\begin{equation}\label{eq:RLS-mean}
    \mathbb{E}\big[\|\Delta\theta_i\|_F\boldsymbol{1}_{\mathrm{PE}_i}\big]\leq\frac{\sqrt{c_x}}{\alpha\sqrt{t_0+i}},\quad
    c_x:=\mathrm{Tr}(\Sigma_w)\big((1+b_K(J_0)^2)\,\bar x(J_0)+\mathrm{Tr}(\Sigma_e)\big),
\end{equation}
with $\bar x(J_0)$ as in \eqref{eq:ms-bound}. If, in addition, Assumption~\ref{as:PE} holds, then for any $\vartheta>0$ and $\delta'\in(0,1-\delta_5)$ with $t_0\geq\frac{8c_x}{\alpha^2\vartheta^2\delta'}$,
\begin{equation}\label{eq:RLS-hp}
    \mathbb{P}\big(\{\|\Delta\theta_i\|_F\leq\vartheta,\ \forall i\in\mathbb{Z}_+\}\cap\mathrm{PE}\big)\geq1-\delta'-\delta_5.
\end{equation}
\end{Theorem}

By Lemma~\ref{lem:indirect-oracle} and Theorem~\ref{thm:RLS}, at any $K\in\mathcal{S}(J_0)$, the bias of the indirect oracle satisfies $\mathbb{E}\big[\|\Delta_I\|_F\boldsymbol{1}_{\{\|\Delta\theta_i\|_F\leq p_\theta\}\cap\mathrm{PE}_i}\big]\leq p\,\mathbb{E}\big[\|\Delta\theta_i\|_F\boldsymbol{1}_{\mathrm{PE}_i}\big]=O(i^{-1/2})$, so it decays as data accumulate, with a single new sample per iteration. In Algorithm~\ref{alg:indirect}, the excitation gains are $\bar K_t=K_0$ for $t<t_0$ and $\bar K_{t_0+i}=K^{\rm ex}_{i+1}$ for $i\in\mathbb{Z}_+$. In the off-policy case, $\bar K_t\equiv K_0\in\mathcal{S}(J_0)$, and the conditions of Theorem~\ref{thm:RLS} hold trivially; moreover, the bound $\bar x(J_0)$ in $c_x$ is conservative: it can be replaced by a bound linear in $C(K_0)$, obtained from the stationary state covariance under $K_0$, which considerably reduces $c_x$ and hence the required $t_0$. In the on-policy case, $\bar K_{t_0+i}=K_{i+1}$: the slow variation of the gains can be ensured by the step size, and the confinement to $\mathcal{S}(J_0)$ with high probability is established in Theorem~\ref{thm:indirect}, where both are made precise.

\subsubsection{Direct Gradient Oracle}\label{sec:direct}
The direct scheme estimates the gradient from empirical cost evaluations, without identifying the system. To this end, we introduce the smoothed cost $C_{v}(K) := \mathbb{E}_{U \sim \mathbb{B}_{v}} \left[ C(K + U) \right], ~K\in\mathcal S,\ v\le h(C(K)),$
where $\mathbb{B}_{v}$ denotes the uniform distribution over all $n_u\times n_x$ matrices with Frobenius norm at most the smoothing radius $v$. It is shown in \cite{pmlr-v80-fazel18a} that the gradient of the smoothed cost satisfies
\begin{equation}\label{zero}
\nabla C_{v}(K) = \mathbb{E}_{U \sim \mathbb{B}_{v}} \left[ \nabla C(K + U) \right]
= \frac{n_x n_u}{v^{2}} \mathbb{E}_{U \sim \mathbb{S}_{v}} \left[ C(K + U) U \right],K\in\mathcal S,v\le h(C(K)),
\end{equation}
where $\mathbb{S}_{v}$ denotes the uniform distribution over the boundary of the Frobenius-norm ball of radius $v$. A rollout of length $\ell$ under a gain $\bar K$ is a trajectory $\{x_t\}_{t=0}^{\ell-1}$ of \eqref{system} under $u_t=\bar Kx_t$, started from $x_0\sim\mathcal{N}(0,\Sigma_0)$. Replacing the expectation in \eqref{zero} by an average over $n$ perturbations and the cost by its empirical value over rollouts of length $\ell$ yields the estimator
\begin{equation}\label{gradientseatimes}
\hat{\nabla}_D C(K,v,\ell,n) = \frac{1}{n} \sum_{k=1}^{n} \frac{n_x n_u}{v^2} \left( \frac{1}{\ell} \sum_{t=0}^{\ell - 1} x_t^{(k)\top} Q_{\bar K_k}\, x_t^{(k)} \right) U_k,
\end{equation}
where $U_k\sim\mathbb{S}_v$ are i.i.d., $\bar K_k:=K+U_k$, $Q_{\bar K_k}:=Q+\bar{K}_k^\top R\bar{K}_k$. 

The resulting scheme is summarized in Algorithm~\ref{Algo1}. At each iteration, the gradient at the current gain $K_i$ is estimated from $n_i$ independent rollouts of length $\ell_i$ with smoothing radius $v_i$, and the gain is updated by a gradient step. The parameters $(v_i,\ell_i,n_i)$ may vary along the iterations to control the bias and the second moment of the estimate. We will specify them later in Theorem~\ref{thm:direct}.
\begin{algorithm}[htb]
\caption{Direct data-driven policy gradient}\label{Algo1}
\begin{algorithmic}
\REQUIRE $K_0\in\mathcal{S}$; step sizes $\{\eta_i\}$; smoothing radii $\{v_i\}$; rollout lengths $\{\ell_i\}$; numbers of rollouts $\{n_i\}$.
\FOR{$i=0,1,\dots$}
\FOR{$k=1,\dots,n_i$}
\STATE Draw $U_k$ uniformly from $\mathbb{S}_{v_i}$ and set $\bar K_k=K_i+U_k$.
\STATE Draw $x_0^{(k)}\sim\mathcal{N}(0,\Sigma_0)$, apply $u_t^{(k)}=\bar K_kx_t^{(k)}$ for $\ell_i$ steps and collect $\{x_t^{(k)}\}_{t=0}^{\ell_i-1}$.
\STATE Compute the empirical cost $\hat C_{\bar K_k}:=\frac{1}{\ell_i}\sum_{t=0}^{\ell_i-1}x_t^{(k)\top}Q_{\bar K_k}x_t^{(k)}$.
\ENDFOR
\STATE Compute $\hat g_i:=\hat\nabla_DC(K_i,v_i,\ell_i,n_i)=\frac{1}{n_i}\sum_{k=1}^{n_i}\frac{n_xn_u}{v_i^2}\hat C_{\bar K_k}U_k$.
\STATE Update $K_{i+1}=K_i-\eta_i\hat g_i$.
\ENDFOR
\end{algorithmic}
\end{algorithm}
We now characterize the oracle \eqref{gradientseatimes}.

\begin{Lemma}[Gradient Oracle from Direct Method]\label{GDDI}
Given $K\in \mathcal{S}$, if the smoothing radius satisfies $v\leq \min \{h(C(K)),\tilde{c}\}$ with $h(\cdot),\tilde{c}$ as in Lemma~\ref{Lipschitz}, then the estimator \eqref{gradientseatimes} satisfies
    \begin{subequations}
        \begin{align}
            \mathbb{E}[\hat{\nabla}_D C(K,v,\ell,n)] &= \nabla C(K) + \Delta_D(K,v,\ell), \\
            \mathbb{E}\big[\lVert \hat{\nabla}_D C(K,v,\ell,n)\rVert_F^2 \big] &\leq V_D(C(K),v,\ell,n),
        \end{align}
    \end{subequations}
    where $\Delta_D(K,v,\ell):=\mathbb{E}[\hat{\nabla}_DC(K,v,\ell,n)-\nabla C(K)]$ satisfies $\lVert \Delta_D(K,v,\ell)\rVert_F \leq\\\bar{\Delta}_D(C(K),v,\ell)$, with $\bar{\Delta}_D$ and $V_D$ defined in \eqref{Deltak} and \eqref{Va2}, and
    \begin{equation}\label{r1}
        \bar{\Delta}_D(C(K),v,\ell)= O\Big(\frac{1}{v\ell}+v\Big),\qquad V_D(C(K),v,\ell,n)= O\Big(1+\frac{1}{\ell^2 v^2}+\frac{1}{nv^2}\Big).
    \end{equation}
\end{Lemma}

Lemma~\ref{GDDI} characterizes the first and second moments of the oracle \eqref{gradientseatimes}. As in \cite{pmlr-v80-fazel18a,Full}, the quality of the estimate is governed by three parameters: the rollout length $\ell$, the smoothing radius $v$, and the number of rollouts $n$. The pair $(\ell,v)$ controls the bias: by \eqref{r1}, the bias is small when $v$ is small and $\ell$ is large. However, a small $v$ inflates the second moment through the terms $\frac{1}{nv^2}$ and $\frac{1}{\ell^2 v^2}$, which can be compensated only by increasing $n$ and $\ell$. Since the rollouts at iteration $i$ of Algorithm~\ref{Algo1} use fresh randomness, independent of the past, Lemma~\ref{GDDI} holds conditionally on the past with $K=K_i$; in contrast to the indirect oracle, the bias is then a deterministic function of $K_i$ and $i$.
\subsection{Closing the Loop between SGD and Gradient Estimators}\label{sec:compare}
We now apply Theorem~\ref{thm:main-convergence} to Algorithms~\ref{alg:indirect} and~\ref{Algo1} by verifying Assumption~\ref{as:oracle} for the two oracles.

\subsubsection{Indirect Method}\label{sec:indirectclosed}
In the off-policy case, the data are generated by the fixed gain $K_0$, independently of the iterates. In the on-policy case, they are generated by the iterates themselves, so the estimation error and the iterates must be controlled jointly. Theorem~\ref{thm:indirect} covers both cases.

\begin{Theorem}\label{thm:indirect}
Consider Algorithm~\ref{alg:indirect}, operated off-policy or on-policy, with $K_0\in\mathcal{S}$, let $J_0>C(K_0)$, and suppose that Assumption~\ref{as:PE} holds. Let $c_x$, $p_\theta$ and $p$ be as in \eqref{eq:RLS-mean}, \eqref{eq:ptheta-uniform} and \eqref{eq:p-def}, define $\Delta_0(J_0):=p\sqrt{c_x}/\alpha$, let $\delta_4\in(0,1-\delta_5)$, and suppose that
\begin{equation}\label{eq:t0-indirect}
    t_0\geq\max\Big\{T_{\rm PE},\ \frac{8\,c_x}{\alpha^2\,(p_\theta)^2\,\delta_4}\Big\}.
\end{equation}
Let the step sizes satisfy $\underline\eta(i+1)^{-\kappa}\le\eta_i\le\bar\eta(i+1)^{-\kappa}$ for all $i\in\mathbb Z_+$, with $\kappa\in(\frac12,1)$ and constants $0<\underline\eta\le\bar\eta<\mu$, and suppose that \eqref{stepsize} in Corollary \ref{cor:event-F} holds with $f=C$, $X_0=K_0$, $c(J_0):=\bar g(J_0)^2$, $\Delta_0(J_0)$ and $\beta=\frac{1}{2}$, and, in the on-policy case, in addition that
\begin{equation}\label{eq:eta-variation}
    \bar\eta\le\frac{\varepsilon_K(J_0)}{\bar g(J_0)}.
\end{equation}
Then there exists an event $E\subseteq F\cap G\cap\mathrm{PE}$ with $\mathbb{P}(E)\geq1-\delta-\delta_4-\delta_5$, where $F$ is the event \eqref{eventeqf} for $\{K_i\}$, with $\|\cdot\|=\|\cdot\|_F$ in \eqref{eventeqf}, $G:=\{\|\Delta\theta_i\|_F\le p_\theta,\ \forall i\in\mathbb{Z}_+\}$, and $\mathrm{PE}$ is the event \eqref{eq:PE-events}.
Moreover, the bias $\Delta_i:=\hat g_i-\nabla C(K_i)$ satisfies
\begin{equation}\label{eq:indirect-bias-rate}
    \mathbb{E}\big[\|\Delta_i\|_F\boldsymbol{1}_E\big]\leq p\,\mathbb{E}\big[\|\Delta\theta_i\|_F\boldsymbol{1}_E\big]\leq\Delta_0(J_0)\,(i+1)^{-1/2},\qquad\forall i\in\mathbb{Z}_+,
\end{equation}
and the following hold:
\begin{enumerate}
\item For every $i\in\mathbb{Z}_+$, $\mathbb{E}\big[(C(K_i)-C(K^*))\boldsymbol{1}_E\big]$ is bounded by the right-hand side of \eqref{finite} in Theorem \ref{thm:main-convergence} with $\mu(J_0)=\mu$, $\beta=\frac{1}{2}$, $c(J_0)=\bar g(J_0)^2$ and $\Delta_0(J_0)$ as above.
\item For any $\lambda\in(\frac{3}{2}-\kappa,1)$, $(C(K_i)-C(K^*))\boldsymbol{1}_E=o\big(i^{-(1-\lambda)}\big)$ almost surely.
\end{enumerate}
\end{Theorem}

The key step of the proof is to verify that the oracle meets the conditions of Section~\ref{sec:Convergence}. By Lemma~\ref{lem:indirect-oracle}, the bias is controlled by the estimation error and the second moment by $\bar g(J_0)^2$. By Theorem~\ref{thm:RLS}, the estimation error decays in expectation at rate $(i+1)^{-1/2}$ on the excitation events $\mathrm{PE}_i$, so the bias vanishes at the same rate on $E$, as in \eqref{eq:indirect-bias-rate}; the persistency of excitation enters only through the probability $\delta_5$ of $E$. Since the estimate is updated recursively, this is achieved with a single new sample per iteration. Since $\beta=\frac{1}{2}<\kappa$, the bound of item~1 is bias-limited and decays as $(i+1)^{-1/2}$ for every $\kappa\in(\frac{1}{2},1)$, so that $\kappa$ only affects its transient; the almost-sure rate of item~2, in contrast, improves as $\kappa$ increases.

\begin{Remark}[Role of the step size]\label{rem:stepsize-indirect}
The step size enters Theorem~\ref{thm:indirect} in two ways. First, as in Theorem~\ref{thm:main-convergence}, condition \eqref{stepsize} keeps the iterates in $\mathcal{S}(J_0)$ and ensures their convergence, in both the off-policy and the on-policy case. Second, in the on-policy case, the iterates also generate the data, so that the closed loop is time-varying. As long as $K_i\in\mathcal{S}(J_0)$ and $\|\Delta\theta_i\|_F\leq p_\theta$, Lemma~\ref{lem:grad-error} gives $\|\hat g_i\|_F\leq\bar g(J_0)$, and \eqref{eq:eta-variation} yields $\|K_{i+1}-K_i\|\leq\bar{\eta}\bar g(J_0)\leq\varepsilon_K(J_0)$. By Lemma~\ref{lem:ms-bound}, this keeps the closed-loop state bounded in mean square without requiring the gains to converge. In the off-policy case, the data are generated by the fixed gain $K_0$, and \eqref{eq:eta-variation} is not needed.
\end{Remark}

\begin{Remark}[Certainty-equivalent control versus policy gradient]\label{rem:CE}
A natural alternative in the indirect setting is the certainty-equivalent controller~\cite{mania2019}, obtained by solving the Riccati equation for the current estimates. In the off-policy case, this is a well-studied approach, and Algorithm~\ref{alg:indirect} mainly serves to illustrate the oracle framework of Section~\ref{sec:Convergence}. In the on-policy case, the data are generated by the time-varying closed loop $x_{t+1}=(A+B\bar{K}_t)x_t+Be_t+w_t$, whose stability does not follow from that of the individual gains: a gain sequence that varies too rapidly can destabilize the loop even if each $A+B\bar{K}_t$ is Schur stable. The certainty-equivalent gain can change abruptly between iterations, whereas the policy gradient increment is bounded by the step size, and \eqref{eq:eta-variation} turns this into the mean-square bound of Lemma~\ref{lem:ms-bound}, in line with~\cite{zhao2025policygradientadaptivecontrol}.
\end{Remark}

\subsubsection{Direct Method}\label{Sec:Direct}
Since the perturbed gains $K_i+U_k$ are applied to the system, the data are always generated on-policy. Let $\mathcal{F}_i$ be the $\sigma$-algebra generated by the rollouts of iterations $0,\dots,i-1$, so that $K_i$ is $\mathcal{F}_i$-measurable. Since the rollouts at iteration $i$ are independent of $\mathcal{F}_i$, Lemma~\ref{GDDI} applies conditionally on $\mathcal{F}_i$ with $K=K_i$; the parameters $(v_i,\ell_i,n_i)$ are then chosen so that the bias vanishes and the second moment remains bounded.

\begin{Theorem}\label{thm:direct}
Consider Algorithm~\ref{Algo1} with $K_0\in\mathcal{S}$, and let $J_0>C(K_0)$. Let the algorithm parameters be
\begin{equation}\label{eq:direct-params}
    v_i=v_0(i+1)^{-1/2},\qquad\ell_i=\lceil\ell_0(i+1)\rceil,\qquad n_i=\lceil n_0(i+1)\rceil,
\end{equation}
with $v_0\leq\min\{h(J_0),\tilde{c}\}$ introduced in Lemma \ref{Lipschitz} and $\ell_0,n_0>0$. Then, for all $i\in\mathbb{Z}_+$ and $K\in\mathcal{S}(J_0)$,
\begin{equation}\label{temp4}
    \bar\Delta_D(C(K),v_i,\ell_i)\leq\Delta_0(J_0)\,(i+1)^{-1/2},\qquad
    V_D(C(K),v_i,\ell_i,n_i)\leq\bar V_D(J_0),
\end{equation}
where $\Delta_0(J_0):=\frac{n_xn_u\,\epsilon''(J_0+v_0h_C(J_0))}{v_0\ell_0}+v_0h_\nabla(J_0)$ with $\epsilon''$ introduced before \eqref{Deltak} and $\bar V_D(J_0):=\sup_iV_D(J_0,v_i,\ell_i,n_i)<\infty$. Hence Assumption~\ref{as:oracle} holds with $c(J_0)=\bar V_D(J_0)$, $\Delta_0(J_0)$ and $\beta=\frac{1}{2}$.

Further, let the step sizes satisfy $\underline\eta(i+1)^{-\kappa}\le\eta_i\le\bar\eta(i+1)^{-\kappa}$ for all $i\in\mathbb Z_+$, with $\kappa\in(\frac12,1)$ and constants $0<\underline\eta\le\bar\eta<\mu$, and suppose that \eqref{stepsize} in Corollary \ref{cor:event-F} holds with $f=C$, $X_0=K_0$, $c(J_0)=\bar V_D(J_0)$, $\Delta_0(J_0)$ and $\beta=\frac{1}{2}$. Then the event $F$ occurs with probability at least $1-\delta$, and the following hold:
\begin{enumerate}
\item For every $i\in\mathbb{Z}_+$, $\mathbb{E}\big[(C(K_i)-C(K^*))\boldsymbol{1}_F\big]$ is bounded by the right-hand side of \eqref{finite} in Theorem \ref{thm:main-convergence} with $\mu(J_0)=\mu$, $\beta=\frac{1}{2}$, $c(J_0)=\bar V_D(J_0)$ and $\Delta_0(J_0)$ as above.
\item For any $\lambda\in(\frac{3}{2}-\kappa,1)$, $(C(K_i)-C(K^*))\boldsymbol{1}_F=o\big(i^{-(1-\lambda)}\big)$ almost surely.
\end{enumerate}
\end{Theorem}

To guarantee convergence to the optimal policy, the smoothing radius $v_i$ must decrease and the rollout length $\ell_i$ must increase, so that the bias vanishes at rate $(i+1)^{-1/2}$. However, a smaller $v_i$ inflates the second moment, which is compensated by increasing the number of rollouts $n_i$ so that $n_iv_i^2$ remains bounded away from zero. As for the indirect method, the bound of item~1 is bias-limited and decays as $(i+1)^{-1/2}$ for every $\kappa\in(\frac{1}{2},1)$, while the almost-sure rate of item~2 improves as $\kappa$ increases. 
\subsection{Comparison between the Two Gradient Estimators}\label{compare2}
From Theorems~\ref{thm:indirect} and~\ref{thm:direct}, both schemes are bias-limited and converge at rate $(i+1)^{-1/2}$ in terms of iterations, but they differ as follows.
\begin{itemize}
    \item \emph{Samples.} The indirect method reuses all past data and collects one new sample per iteration, so that after $s= t_0+i$ samples the expected optimality gap is $O(s^{-1/2})$. The direct method cannot reuse data: to keep the second moment bounded while $v_i\to0$, it requires $n_i\ell_i$ of order $i^2$ new samples at iteration $i$, i.e., $s=O(i^3)$ in total, and the gap is $O(s^{-1/6})$.
    \item \emph{Computation.} Per iteration, the indirect method performs a rank-one RLS update and solves two Lyapunov equations of dimension $n_x$; the direct method uses $n_i\ell_i$, of order $i^2$, time steps of interaction but solves no matrix equation.
    \item \emph{Data and excitation.} The indirect method uses a single data stream with dithering, can be operated off-policy or on-policy, relies on the persistency of excitation in Assumption~\ref{as:PE}, and requires an offline phase of $t_0$ samples; together, these cost the additional probability $\delta_4+\delta_5$. The direct method uses independent rollouts, is inherently on-policy since it perturbs the applied gain, and requires neither an excitation assumption nor an offline phase.
\end{itemize}

\section{Simulation Results}
In this section, we present numerical results\footnote{The MATLAB code used to generate these results is available at \url{https://github.com/col-tasas/2026-SGDLQR}.} that illustrate the theoretical findings of the previous sections. Throughout, we consider the linearized longitudinal dynamics of a Boeing 747 aircraft~\cite{doi:10.1137/23M1554771}, discretized using a zero-order hold with a sampling time of $10\,\mathrm{s}$,
\vspace{-10pt}
\begin{align}\label{LTIsimulation}
x_{t+1}=Ax_t&+Bu_t+w_t,\\ \nonumber
A = 
\begin{bmatrix}
1 & -1.13 & -0.65 & -8.07 & 1.59 \\
0 & 0.77 & 0.32 & -0.98 & -2.97 \\
0 & 0.12 & 0.02 & 0 & -0.36 \\
0 & 0.01 & 0.01 & -0.03 & -0.04 \\
0 & 0.14 & -0.09 & 0.29 & 0.76
\end{bmatrix},
&B =
\begin{bmatrix}
89.20 & -50.17 & 1.13 & -19.35 \\
5.22 & 6.36 & 0.23 & -0.32 \\
-9.47 & 5.93 & -0.12 & 0.99 \\
-0.32 & 0.32 & -0.01 & -0.01 \\
-4.53 & 3.21 & -0.14 & 0.09
\end{bmatrix}
\end{align}
The weight matrices are $Q=I_5$ and $R=I_4$, the initial state is $x_0\sim\mathcal{N}(0,10^{-6}I_5)$, and $w_t\sim\mathcal{N}(0,10^{-3}I_5)$. The initial gain $K_0$ is the optimal LQR gain for $(A,B,40Q,R)$. The performance is measured by the relative cost gap $\frac{C(K_i)-C(K^*)}{C(K^*)}$. In all Monte Carlo studies, a run contributes to the mean up to the last iteration before $K_i$ becomes destabilizing and is discarded afterwards; the number of such runs is reported in the captions.
\subsection{Convergence of SGD with Biased Gradients}
We first illustrate the role of a vanishing step size and a vanishing bias. The SGD iteration \eqref{eq:sgd-iteration} is run with the oracle $\hat g_i=\nabla C(K_i)+\Delta_i+W_i,~ \Delta_i:=\bar\Delta_i\,D,$ where $D$ is a direction with $\|D\|_F=1$, drawn at random once per run, and $W_i$ is zero-mean Gaussian noise with $\mathbb{E}\|W_i\|_F^2=5\times10^{-5}$, independent across iterations. Given the past, $\Delta_i$ is the bias of this oracle, with $\|\Delta_i\|_F=\bar\Delta_i$. We compare a constant step size $\eta_i=0.05$ with the decreasing step size $\eta_i=0.05\,(i_\eta/\max\{i,i_\eta\})^{0.51}=O(i^{-0.51})$, and a constant bias $\bar\Delta_i=0.2$ with the vanishing bias $\bar\Delta_i=0.2\,(i_\Delta/\max\{i,i_\Delta\})^{1/2}=O(i^{-1/2})$, which satisfies Assumption~\ref{as:oracle} with $\beta=\frac12$, where $i_\eta=i_\Delta=5$. Each case is run for $10^7$ iterations.
\begin{figure}[H]
    \centering
    \includegraphics[width=0.9\linewidth]{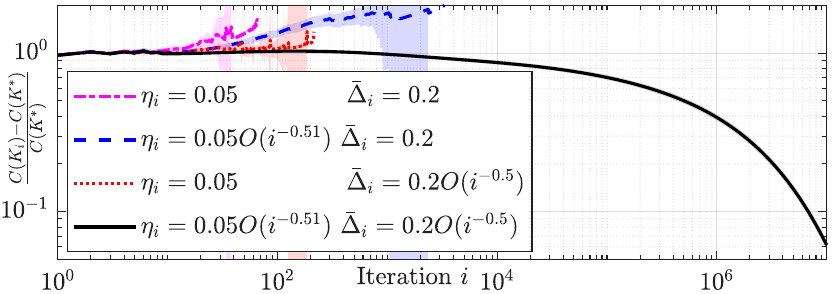}
\caption{SGD with different step sizes and biases: mean relative cost gap over 100 Monte Carlo runs with 99.9\% confidence intervals. The bias direction is drawn at random once per run. Number of diverged runs: constant step size and constant bias (magenta dash-dotted), 100; decreasing step size and constant bias (blue dashed), 100; constant step size and vanishing bias (red dotted), 100; decreasing step size and vanishing bias (black solid), 19. 
}
    \label{fig:1}
\end{figure}
If the bias does not vanish, the runs diverge, irrespective of whether the step size is constant (magenta dash-dotted line) or decreasing (blue dashed line). If the bias vanishes but the step size is kept constant (red dotted line), the runs also diverge. In all three cases, $\sum_i\eta_i\bar\Delta_i=\infty$, and additionally $\sum_i\eta_i^2=\infty$ for the magenta and red lines, so that \eqref{biasterm} fails and Theorem~\ref{thm:main-convergence} does not apply. In particular, the blue dashed line, for which $\sum_i\eta_i^2<\infty$, illustrates that a vanishing step size does not suffice when the accumulated bias is not summable, and the red dotted line illustrates that a vanishing bias alone does not suffice either. Only when both the bias and the step size vanish (black solid line) does the cost converge to the optimum, consistent with Theorem~\ref{thm:main-convergence}. The $19$ diverged runs in this case leave the stabilizing set early on, i.e., they lie outside the event $F$ of Theorem~\ref{thm:main-convergence}.
\subsection{Data-driven Policy Gradient Methods}
\begin{figure}[H]
    \centering
    \includegraphics[width=0.9\linewidth]{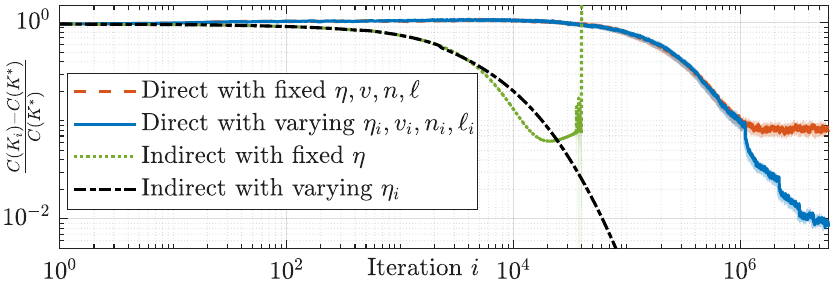}
\caption{Indirect and direct data-driven policy gradient methods: mean relative cost gap over 100 Monte Carlo runs with 99.9\% confidence intervals. Number of diverged runs: indirect method with fixed $\eta$ (green dotted), 100; with varying $\eta_i$ (black dash-dotted), 0; direct method with fixed $\eta,v,n,\ell$ (red dashed), 4; with varying $\eta_i,v_i,n_i,\ell_i$ (blue solid), 5.}
    \label{fig:4}
\end{figure}

Finally, we compare the indirect and the direct data-driven policy gradient methods, each with constant and with time-varying parameters. Fig.~\ref{fig:4} shows the mean relative cost gap over $100$ Monte Carlo runs per method together with $99.9\%$ confidence intervals of the mean. For the indirect method, operated on-policy, the dithering signal is $e_t\sim\mathcal{N}(0,I_4)$ and $t_0=50$. We compare the constant step size $\eta=0.05$ with the decreasing step size $\eta_i=\eta\,(i_\eta/\max\{i,i_\eta\})^{0.51}=O(i^{-0.51})$ with $i_\eta=2000$, which is of the form required in Theorem~\ref{thm:indirect}. For the direct method, we compare the constant parameters of~\cite{pmlr-v80-fazel18a,Full}, $n=100$, $\ell=10$, $v=0.01$, $\eta=0.002$, with the time-varying parameters $n_i=n\lceil \frac{i+1}{n_c}\rceil=O(i)$, $\ell_i=\ell\lceil \frac{i+1}{n_c}\rceil=O(i)$, $v_i=v\,(i_v/\max\{i,i_v\})^{1/2}=O(i^{-1/2})$ and $\eta_i=\eta\,(i_\eta/\max\{i,i_\eta\})^{0.51}=O(i^{-0.51})$, with $n_c=1.1\times10^6$, $i_v=3.3\times10^6$ and $i_\eta=10^6$. The two parameter sets coincide up to iteration $10^6$; over the plotted horizon of $5\times10^6$ iterations, $\eta_i$ decreases by a factor of about $2.3$, $n_i$ and $\ell_i$ increase by a factor of $5$, and $v_i$ decreases by about $20\%$.
With the constant step size $\eta=0.05$, the indirect method (green dotted) decreases the cost over a long horizon, but all runs eventually diverge. With a constant step size, the accumulated effect of the random bias, $\sum_i\eta_i\Delta_0(J_0)(i+1)^{-1/2}$, is unbounded, so that \eqref{stepsize} fails and Theorem \ref{thm:indirect} does not apply. For the direct method with constant parameters (red dashed), the cost stalls at a relative gap of approximately $10\%$, since the gradient estimation error does not vanish. With time-varying parameters (blue solid), the two curves coincide up to iteration $10^6$; afterwards, the decreasing step size and the increasing rollout length, which reduces the bias term $O(1/(v\ell))$ in \eqref{r1}, let the cost continue to decrease below this level.

Both methods require an unbounded number of samples to reach the exact optimum. To reach a relative cost gap of $1\%$, the indirect method with decreasing step size requires about $8\times10^4$ samples, whereas the direct method requires about $4.8\times10^{10}$ samples with time-varying parameters. This is consistent with the comparison in Section~\ref{compare2}: the indirect method reuses all past data, whereas the direct method draws fresh rollouts at each iteration.

\section{Conclusion}\label{sec:conclusion}
We studied SGD with biased stochastic gradient oracles for objectives that are only locally smooth and locally gradient dominated, with constants that may grow with the objective value. We showed that, if the bias vanishes sufficiently fast in expectation and the second moment of the oracle remains bounded on a sublevel set, then, on an event of prescribed probability, the optimality gap vanishes almost surely with an asymptotic rate and in expectation with an explicit bound at every iteration. We applied this framework to data-driven policy gradient for the LQR, with an indirect estimator based on recursive least squares, operated off-policy or on-policy, and a direct zeroth-order estimator. This yields convergence guarantees for all iterations for both schemes and shows that the indirect estimator, by reusing past data, attains the same rate in iterations with substantially fewer samples. Limitations include the dependence of the conditions on unknown system constants and the persistency of excitation assumption for the indirect estimator. Future work includes extending the framework to accelerated and constrained methods.

\section*{Acknowledgments}
Bowen Song acknowledges the support of the International Max Planck Research School for Intelligent Systems (IMPRS-IS).

\appendix
\section{Proofs in Section \ref{sec:Convergence}}
\subsection{Proof of Lemma \ref{lem:event-C}}\label{ProofLemma4}
\begin{proof}
Fix $k\in\mathbb{Z}_{++}$. By \eqref{eq:sgd-iteration},
$\|X_k-X_{k-1}\|=\eta_{k-1}\|\hat\nabla f(X_{k-1},\xi_{k-1})\|$, hence
\vspace{-5pt}
\begin{equation*}
    F_{k-1}\cap\{\|X_k-X_{k-1}\|>r(J_0)\}
    =\Big\{\boldsymbol{1}_{F_{k-1}}\|\hat\nabla f(X_{k-1},\xi_{k-1})\|^2>\tfrac{r(J_0)^2}{\eta_{k-1}^2}\Big\}.
\end{equation*}
By Markov's inequality,
\vspace{-10pt}
\begin{equation*}
\begin{split}
    \mathbb{P}\big(F_{k-1}&\cap\{\|X_k-X_{k-1}\|>r(J_0)\}\big)
    \leq \frac{\eta_{k-1}^2}{r(J_0)^2}\,
      \mathbb{E}\big[\boldsymbol{1}_{F_{k-1}}\|\hat\nabla f(X_{k-1},\xi_{k-1})\|^2\big]\\
    &= \frac{\eta_{k-1}^2}{r(J_0)^2}\,
      \mathbb{E}\Big[\boldsymbol{1}_{F_{k-1}}\,
      \mathbb{E}\big[\|\hat\nabla f(X_{k-1},\xi_{k-1})\|^2\,\big|\,\mathcal{F}_{k-1}\big]\Big]
    \leq \frac{c(J_0)\,\eta_{k-1}^2}{r(J_0)^2},
\end{split}
\end{equation*}
where the equality holds because $F_{k-1}$ is $\mathcal{F}_{k-1}$-measurable, and the
last inequality uses $X_{k-1}\in \mathcal{S}(J_0)$ on $F_{k-1}$ and Assumption~\ref{as:oracle}.
A union bound over $k\in\mathbb{Z}_{++}$ and \eqref{eq:lemC-condition} yield $\mathbb{P}\Big(\bigcup_{k\in\mathbb{Z}_{++}} F_{k-1}\cap\{\|X_k-X_{k-1}\|>r(J_0)\}\Big)
    \leq \frac{c(J_0)}{r(J_0)^2}\sum_{k=1}^{\infty}\eta_{k-1}^2\leq\delta. $
\end{proof}
\subsection{Proof of Lemma~\ref{lem:event-F}}
\label{ProofLemmaF}
\begin{proof}
For brevity, write $\hat g_i:=\hat\nabla f(X_i,\xi_i)$, $D_i:=f(X_i)-f(X^*)$, $B_i:=\{\|X_{i+1}-X_i\|\leq r(J_0)\}$, and $ \zeta_i:=-\langle\hat g_i-\nabla f(X_i),\nabla f(X_i)\rangle$, 
\vspace{-10pt}
\begin{equation*}
M_i:=\sum_{k=0}^{i}\eta_k\boldsymbol{1}_{F_k}\zeta_k,\quad
    S_i:=\frac{L(J_0)}{2}\sum_{k=0}^{i}\eta_k^2\boldsymbol{1}_{F_k}\|\hat g_k\|^2,\quad
    R_i:=M_i^2+S_i,
\end{equation*}
with $M_{-1}=S_{-1}=R_{-1}:=0$.

\emph{Step 1 (descent on the confinement event).} Let $k\in\mathbb{Z}_+$ and suppose that $F_k\cap B_k$ occurs. Then $f(X_k)\leq J_0$, so $\|X_{k+1}-X_k\|\leq r(J_0)\leq r(f(X_k))$ and $L(f(X_k))\leq L(J_0)$ by the monotonicity of $r$ and $L$. Definition~\ref{def:local-smooth} with $X_{k+1}-X_k=-\eta_k\hat g_k$ gives
\begin{equation}\label{eq:pf-F-descent}
    D_{k+1}\leq D_k-\eta_k\|\nabla f(X_k)\|^2+\eta_k\zeta_k+\frac{L(J_0)}{2}\eta_k^2\|\hat g_k\|^2
    \leq D_k+\eta_k\zeta_k+\frac{L(J_0)}{2}\eta_k^2\|\hat g_k\|^2.
\end{equation}

\emph{Step 2 (reduction to two events).} Define
\vspace{-5pt}
\begin{equation*}
    \Gamma:=\Big(\bigcup_{k\in\mathbb{Z}_{++}}F_{k-1}\cap\{\|X_k-X_{k-1}\|>r(J_0)\}\Big)^{c},\qquad
    \varepsilon_R:=\{R_i\leq\epsilon,\ \forall i\in\mathbb{Z}_+\}.
\end{equation*}
We show $\Gamma\cap \varepsilon_R\subseteq F$. Fix an outcome in $\Gamma\cap \varepsilon_R $. The event $F_0$ occurs since $f(X_0)<J_0$. Suppose that $F_i$ occurs. Then $F_k$ occurs for all $k\leq i$, and since the outcome lies in $\Gamma$, so does $B_k$. Hence \eqref{eq:pf-F-descent} holds for $k=0,\dots,i$ with $\boldsymbol{1}_{F_k}=1$, and summing yields
\vspace{-5pt}
\begin{equation*}
    D_{i+1}\leq D_0+M_i+S_i\leq D_0+\sqrt{R_i}+R_i\leq D_0+\sqrt{\epsilon}+\epsilon=J_0-f(X^*).
\end{equation*}
Thus $f(X_{i+1})\leq J_0$, which together with $B_i$ gives $F_{i+1}$. By induction, $F_i$ occurs for all $i\in\mathbb{Z}_+$, i.e., the outcome lies in $F$. By Lemma~\ref{lem:event-C} with $\delta=\delta_3$, $\mathbb{P}(\Gamma^c)\leq\delta_3$; it remains to show $\mathbb{P}(\varepsilon_R^c)\leq\delta_1+\delta_2$.

\emph{Step 3 (conditional moments).} Fix $i\in\mathbb{Z}_+$. The event $F_i$ is $\mathcal{F}_i$-measurable and $X_i\in\mathcal{S}(J_0)$ on $F_i$. Hence, on $F_i$, Assumptions~\ref{as:grad-bound} and \ref{as:oracle} give
\vspace{-5pt}
\begin{subequations}
    \begin{align}
    \big|\mathbb{E}[\zeta_i\mid\mathcal{F}_i]\big| &= \big|\langle\Delta_i,\nabla f(X_i)\rangle\big|\leq b_\nabla(J_0)\,\|\Delta_i\|,\label{eq:pf-F-m1}\\
    \mathbb{E}[\zeta_i^2\mid\mathcal{F}_i] &\leq b_\nabla(J_0)^2\,\mathbb{E}\big[\|\hat g_i-\nabla f(X_i)\|^2\mid\mathcal{F}_i\big]\leq \sigma_\zeta^2(J_0),\label{eq:pf-F-m2}\\
    \mathbb{E}[\|\hat g_i\|^2\mid\mathcal{F}_i] &\leq c(J_0),\label{eq:pf-F-m3}
\end{align}
\end{subequations}
where \eqref{eq:pf-F-m2} follows from the Cauchy--Schwarz inequality and
$\mathbb{E}[\|\hat g_i-\nabla f(X_i)\|^2\mid\mathcal{F}_i]=\mathbb{E}[\|\hat g_i\|^2\mid\mathcal{F}_i]-2\,\langle\mathbb{E}[\hat g_i\mid\mathcal{F}_i],\nabla f(X_i)\rangle+\|\nabla f(X_i)\|^2\leq\big(\sqrt{c(J_0)}+b_\nabla(J_0)\big)^2$, using $\|\mathbb{E}[\hat g_i\mid\mathcal{F}_i]\|\leq\sqrt{c(J_0)}$ by Jensen's inequality.

\emph{Step 4 (probability of $\varepsilon_R $).} Let $\tau$ be the first iteration at which $R_i$ exceeds $\epsilon$, i.e., $\tau:=\inf\{i\in\mathbb{Z}_+: R_i>\epsilon\}$, with $\tau:=\infty$ if $R_i\leq\epsilon$ for all $i$. Then $\varepsilon_R =\{\tau=\infty\}$ and $\varepsilon_R^c=\{\tau<\infty\}$. Since $R_k$ is $\mathcal{F}_{k+1}$-measurable, $\{\tau\geq i\}=\{R_k\leq\epsilon,\ k=0,\dots,i-1\}\in\mathcal{F}_i$, and $|M_{i-1}|\leq\sqrt{R_{i-1}}\leq\sqrt{\epsilon}$ on this event. Since $M_i=M_{i-1}+\eta_i\boldsymbol{1}_{F_i}\zeta_i$, $ R_i-R_{i-1}=\eta_i^2\boldsymbol{1}_{F_i}\zeta_i^2+2\eta_i\boldsymbol{1}_{F_i}\zeta_i M_{i-1}+\frac{L(J_0)}{2}\eta_i^2\boldsymbol{1}_{F_i}\|\hat g_i\|^2.$
Conditioning on $\mathcal{F}_i$ and using \eqref{eq:pf-F-m1}--\eqref{eq:pf-F-m3}, we obtain
\begin{equation}\label{eq:pf-F-incr}
    \mathbb{E}\big[(R_i-R_{i-1})\boldsymbol{1}_{\{\tau\geq i\}}\big]\leq \eta_i^2\Big(\sigma_\zeta^2(J_0)+\frac{L(J_0)c(J_0)}{2}\Big)+2\sqrt{\epsilon}\,b_\nabla(J_0)\,\eta_i\,\mathbb{E}\big[\|\Delta_i\|\boldsymbol{1}_{F_i}\big].
\end{equation}
Since $F_i\subseteq\{X_i\in\mathcal{S}(J_0)\}$, \eqref{eq:bias-decay} gives $\mathbb{E}[\|\Delta_i\|\boldsymbol{1}_{F_i}]\leq\Delta_0(J_0)(i+1)^{-\beta}$. For $n\in\mathbb{Z}_+$, $R_{\min(n,\tau)}=\sum_{i=0}^{n}(R_i-R_{i-1})\boldsymbol{1}_{\{\tau\geq i\}}$. Hence, by \eqref{eq:pf-F-incr} and conditions 1--2,
\vspace{-5pt}
\begin{equation*}
\begin{split}
    \mathbb{E}[R_{\min(n,\tau)}]&\leq \Big(\sigma_\zeta^2(J_0)+\frac{L(J_0)c(J_0)}{2}\Big)\sum_{i=0}^{\infty}\eta_i^2+2\sqrt{\epsilon}\,b_\nabla(J_0)\sum_{i=0}^{\infty}\eta_i\Delta_0(J_0)(i+1)^{-\beta}\\
    &\leq(\delta_1+\delta_2)\,\epsilon.
\end{split}
\end{equation*}
Since $R_{\min(n,\tau)}\geq0$ and $R_{\min(n,\tau)}=R_\tau>\epsilon$ on $\{\tau\leq n\}$, Markov's inequality gives $\mathbb{P}(\tau\leq n)\leq\mathbb{E}[R_{\min(n,\tau)}]/\epsilon\leq\delta_1+\delta_2$. Letting $n\to\infty$ yields $\mathbb{P}(\varepsilon_R ^c)=\mathbb{P}(\tau<\infty)\leq\delta_1+\delta_2$.
\emph{Conclusion.} By Step 2, $\mathbb{P}(F^c)\leq\mathbb{P}(\Gamma^c)+\mathbb{P}(\varepsilon_R ^c)\leq\delta_1+\delta_2+\delta_3=\delta$.
\end{proof}
\section{Proofs in Section \ref{sec:LQR-example}}
\subsection{Proof of Lemma \ref{lem:grad-error}}
\begin{proof}
Let $K\in\mathcal{S}(J_0)$, $A_K:=A+BK$, $\hat A_K:=\hat A+\hat BK$, and write $s_\Sigma:=J_0/\lambda_1(Q)$, $s_P:=J_0/\lambda_1(\Sigma_w)$, $b_K:=b_K(J_0)$ and $a:=\|A\|+\|B\|b_K$, so that $\lVert\Sigma_K\rVert\le s_\Sigma$, $\lVert P_K\rVert\le s_P$, $\|K\|\leq b_K$ and $\|A_K\|\leq a$. We use the perturbation bound for Lyapunov equations of~\cite[Lemma~11]{zhao2025policygradientadaptivecontrol}, stated there for $X=I+A_1XA_1^\top$. For $X=W+A_1XA_1^\top$ with $W\succ0$, combining it with $X\succeq\lambda_1(W)X_I$, where $X_I$ denotes the solution with $W=I$, shows that $\|A_2-A_1\|\leq\lambda_1(W)/(4\|X(A_1)\|(1+\|A_1\|))$ implies that $A_2$ is Schur stable and $\|X(A_2)-X(A_1)\|\leq4\|X(A_1)\|^2(1+\|A_1\|)\|A_2-A_1\|/\lambda_1(W)$. For $\|\Delta\theta\|_F\leq p_\theta$, we have $\|\hat A_K-A_K\|\leq(1+b_K)\|\Delta\theta\|\leq\lambda_1(Q)\lambda_1(\Sigma_w)/(4J_0(1+a))$, which is below this threshold both for $\Sigma_K$ ($W=\Sigma_w$) and for $P_K$ ($W=Q_K\succeq Q$). Hence $\hat A_K$ is Schur stable, $\lVert\hat\Sigma_K-\Sigma_K\rVert\leq4s_\Sigma^2(1+a)(1+b_K)\lVert\Delta\theta\rVert/\lambda_1(\Sigma_w)$, and $\lVert\hat P_K-P_K\rVert\leq4s_P^2(1+a)(1+b_K)\lVert\Delta\theta\rVert/\lambda_1(Q)$. Write $\hat\nabla_IC(K,\hat A,\hat B)-\nabla C(K)=2E_K(\hat\Sigma_K-\Sigma_K)+2(\hat E_K-E_K)\hat\Sigma_K$, and use $\lVert E_K\rVert_F\leq b_E(J_0)$ (\cite[Lemma~11]{pmlr-v80-fazel18a}, with $b_E$ in \eqref{boundedEK}) together with the identity $\hat B^\top\hat P_K-B^\top P_K=\hat B^\top(\hat P_K-P_K)+(\hat B-B)^\top P_K$. This yields $\lVert2E_K(\hat\Sigma_K-\Sigma_K)\rVert\leq p_1\lVert\Delta\theta\rVert$, $\lVert\hat E_K-E_K\rVert\leq(p_2+p_3)\lVert\Delta\theta\rVert$ and $\lVert\hat\Sigma_K\rVert\leq p_4$, where $p_1:=8b_E(J_0)s_\Sigma^2(1+a)(1+b_K)/\lambda_1(\Sigma_w)$; $p_2:=\big[s_P+4(\|B\|+p_\theta)s_P^2(1+a)(1+b_K)/\lambda_1(Q)\big]\big(a+(1+b_K)p_\theta\big)$; $p_3:=\|B\|s_P(1+b_K)$; and $p_4:=s_\Sigma+4s_\Sigma^2(1+a)(1+b_K)p_\theta/\lambda_1(\Sigma_w)$. Since the gradient error has rank at most $\min\{n_x,n_u\}$, combining these bounds gives the first bound in \eqref{eq:grad-error} with
\vspace{-5pt}
\begin{equation}\label{eq:p-def}
    p:=\sqrt{\min\{n_x,n_u\}}\,\big(p_1+2p_4(p_2+p_3)\big).
\end{equation}
The second bound follows from the first, $\|\Delta\theta\|_F\leq p_\theta$, and $\|\nabla C(K)\|_F\leq b_\nabla(J_0)$.
\end{proof}

\subsection{Proof of Lemma~\ref{lem:ms-bound}}\label{app:ms-bound}
\begin{proof}
The argument follows the sequential stability analysis of~\cite[Definition~2, Lemmas~16 and~18]{zhao2025policygradientadaptivecontrol}, adapted to the noise covariance $\Sigma_w$ and to bounds in mean square. Write $A_t:=A+B\bar K_t$, let $\bar\Sigma_t:=\Sigma_{\bar K_t}$ be the solution of $\bar\Sigma_t=\Sigma_w+A_t\bar\Sigma_tA_t^\top$, so that $C(\bar K_t)=\mathrm{Tr}(Q_{\bar K_t}\bar\Sigma_t)$, and abbreviate $\chi:=\chi(J_0)$ and $\varrho:=\varrho(J_0)$.

\emph{Step 1 (strong stability).} Let $\Psi_t:=(\bar\Sigma_t/\lambda_1(\Sigma_w))^{1/2}$ and $\Lambda_t:=\Psi_t^{-1}A_t\Psi_t$, so that $A_t=\Psi_t\Lambda_t\Psi_t^{-1}$. Since $\bar\Sigma_t\succeq\Sigma_w$, we have $\|\Psi_t^{-1}\|\leq1$. Since $C(\bar K_t)\geq\lambda_1(Q)\,\mathrm{Tr}(\bar\Sigma_t)\geq\lambda_1(Q)\|\bar\Sigma_t\|$ and $\bar K_t\in\mathcal{S}(J_0)$, $\|\Psi_t\|^2=\frac{\|\bar\Sigma_t\|}{\lambda_1(\Sigma_w)}\leq\chi^2.$

In particular, $\chi\geq1$, since $\|\bar\Sigma_t\|\geq\lambda_1(\Sigma_w)$. Moreover, $\Lambda_t\Lambda_t^\top=\bar\Sigma_t^{-\frac12}A_t\bar\Sigma_tA_t^\top\bar\Sigma_t^{-\frac12}\\=I-\bar\Sigma_t^{-\frac12}\Sigma_w\bar\Sigma_t^{-\frac12}$, hence $\|\Lambda_t\|^2\leq1-\frac{\lambda_1(\Sigma_w)}{\|\bar\Sigma_t\|}\leq1-\chi^{-2}$, i.e., $\|\Lambda_t\|\leq1-\varrho.$

\emph{Step 2 (slow variation).} Since $\|\Psi_{t+1}^{-1}\Psi_t\|^2=\lambda_{\max}\big(\bar\Sigma_{t+1}^{-1/2}\bar\Sigma_t\bar\Sigma_{t+1}^{-1/2}\big)\leq1+\|\bar\Sigma_t-\bar\Sigma_{t+1}\|/\lambda_1(\Sigma_w)$ and $\sqrt{1+x}\leq1+x/2$ for $x\geq0$, $ \|\Psi_{t+1}^{-1}\Psi_t\|\leq1+\frac{\|\bar\Sigma_{t+1}-\bar\Sigma_t\|}{2\lambda_1(\Sigma_w)}\leq1+\frac{h_\Sigma(J_0)\,\|\bar K_{t+1}-\bar K_t\|}{2\lambda_1(\Sigma_w)}\leq1+\frac{\varrho}{2},$
where the second inequality follows from Lemma~\ref{Lipschitz}, which applies since $\bar K_t,\bar K_{t+1}\in\mathcal{S}(J_0)$ and $\|\bar K_{t+1}-\bar K_t\|\leq h(J_0)\leq h(C(\bar K_t))$, together with $h_\Sigma(C(\bar K_t))\leq h_\Sigma(J_0)$, by the monotonicity of $h$ and $h_\Sigma$. The last inequality follows from $\|\bar K_{t+1}-\bar K_t\|\leq\varepsilon_K(J_0)$.

\emph{Step 3 (transition matrix).} For $t>s$, let $\Phi(t,s):=A_{t-1}\cdots A_s$, and $\Phi(s,s):=I$. Writing $\Phi(t,s)=\Psi_{t-1}\Lambda_{t-1}(\Psi_{t-1}^{-1}\Psi_{t-2})\Lambda_{t-2}\cdots(\Psi_{s+1}^{-1}\Psi_s)\Lambda_s\Psi_s^{-1}$, Steps~1--2 give $  \|\Phi(t,s)\|\leq\chi(1-\varrho)^{t-s}\Big(1+\frac{\varrho}{2}\Big)^{t-s-1}\leq\chi\Big(1-\frac{\varrho}{2}\Big)^{t-s},$
where we used $(1-\varrho)(1+\varrho/2)\leq1-\varrho/2$. Since $\chi\geq1$, the bound also holds for $t=s$. It holds surely, since Steps~1--2 hold for every realization of $\{\bar K_t\}$.

\emph{Step 4 (mean square).} For $t\in\mathbb{Z}_+$, $x_t=\Phi(t,0)\,x_0+\sum_{s=0}^{t-1}\Phi(t,s+1)\,(Be_s+w_s)$. By Minkowski's inequality in $L^2$, Step~3, $\mathbb{E}\|x_0\|^2=\mathrm{Tr}(\Sigma_0)$ and $\mathbb{E}\|Be_s+w_s\|^2=\mathrm{Tr}(B\Sigma_eB^\top)+\mathrm{Tr}(\Sigma_w)$, $ \mathbb{E}\big[\|x_t\|^2\big]^{1/2}\leq\chi\sqrt{\mathrm{Tr}(\Sigma_0)}+\chi\sqrt{\mathrm{Tr}(\Sigma_w)+\mathrm{Tr}(B\Sigma_eB^\top)}\,\sum_{s=0}^{t-1}\Big(1\\-\frac{\varrho}{2}\Big)^{t-s-1},$ and the geometric sum is bounded by $2/\varrho$, which yields \eqref{eq:ms-bound}.
\end{proof}
\subsection{Proof of Theorem~\ref{thm:RLS}}
\begin{proof}
Let $R_i:=\sum_{t=0}^{t_0+i-1}w_td_t^\top$, so that $\Delta\theta_i=R_iH_i^{-1}$ by \eqref{eq:LS} whenever $H_i\succ0$. On $\mathrm{PE}_i$, $H_i\succeq\alpha(t_0+i)I$, hence
\vspace{-10pt}
\begin{equation}\label{eq:Hinv}
    \|\Delta\theta_i\|_F\boldsymbol{1}_{\mathrm{PE}_i}\leq\frac{\|R_i\|_F}{\alpha(t_0+i)}.
\end{equation}
Since $\bar K_t$ is independent of $\{(w_s,e_s)\}_{s\geq t}$, $w_t$ is zero-mean and independent of $d_t$ and of the past. Hence $\{R_i\}$ is a matrix martingale, $\{\|R_i\|_F^2\}$ is a submartingale, and all cross terms vanish, so that $\mathbb{E}\|R_i\|_F^2=\mathrm{Tr}(\Sigma_w)\sum_{t=0}^{t_0+i-1}\mathbb{E}\|d_t\|^2\leq c_x(t_0+i)$, where we used $\mathbb{E}\|d_t\|^2=\mathbb{E}\|x_t\|^2+\mathbb{E}\|\bar K_tx_t\|^2+\mathrm{Tr}(\Sigma_e)\leq(1+b_K(J_0)^2)\bar x(J_0)+\mathrm{Tr}(\Sigma_e)$ by Lemma~\ref{lem:ms-bound} and $\|\bar K_t\|\leq b_K(J_0)$. Combining with \eqref{eq:Hinv} and Jensen's inequality yields \eqref{eq:RLS-mean}.

For the second claim, let $j_0:=\lfloor\log_2t_0\rfloor$ and partition $\mathbb{Z}_+$ into the blocks $\mathcal{B}_j:=\{i\in\mathbb{Z}_+:2^j\leq t_0+i<2^{j+1}\}$, $j\geq j_0$. By \eqref{eq:Hinv} and Doob's maximal inequality applied to $\{\|R_i\|_F^2\}$,
\vspace{-5pt}
\begin{equation*}
    \mathbb{P}\Big(\Big\{\max_{i\in\mathcal{B}_j}\|\Delta\theta_i\|_F>\vartheta\Big\}\cap\mathrm{PE}\Big)\leq\mathbb{P}\Big(\max_{t_0+i<2^{j+1}}\|R_i\|_F\geq\vartheta\alpha2^j\Big)\leq\frac{c_x2^{j+1}}{\vartheta^2\alpha^2\,2^{2j}}=\frac{2c_x}{\vartheta^2\alpha^2}2^{-j}.
\end{equation*}
Summing over $j\geq j_0$ and using $2^{j_0}>t_0/2$ yields $\mathbb{P}(\{\exists i:\|\Delta\theta_i\|_F>\vartheta\}\cap\mathrm{PE})<\frac{8c_x}{\vartheta^2\alpha^2t_0}\leq\delta'$. Since $\mathbb{P}(\mathrm{PE}^c)\leq\delta_5$ by Assumption~\ref{as:PE}, \eqref{eq:RLS-hp} follows.
\end{proof}
\subsection{Proof of Lemma \ref{GDDI}}\label{ProofGDDI}
\begin{proof}
We quantify the errors introduced by the smoothing radius $v$ and the finite rollout length $\ell$. Write $\mathbb{E}[\hat{\nabla}_DC(K,v,\ell,n)]-\nabla C(K)
    ={\nabla C_{v}(K)-\nabla C(K)}
    +{\mathbb{E}[\hat{\nabla}_DC(K,v,\ell,n)]-\nabla C_{v}(K)}.$
Since $\lVert U\rVert_F\leq v\leq \min\{h(C(K)),\tilde c\}$, \eqref{zero} and the Lipschitz continuity of $\nabla C$ (Lemma~\ref{Lipschitz}) give $\lVert\nabla C_v(K)-\nabla C(K)\rVert_F\leq v\,h_\nabla(C(K))$. For the second term, let $C^{(\ell)}(K):=\mathbb{E}_{x_0,w_t}\big[\frac{1}{\ell}\sum_{t=0}^{\ell-1}x_t^\top(Q+K^\top RK)x_t\big]$.

By \cite[Lemma C.1]{Full}, $\lvert C^{(\ell)}(K')-C(K')\rvert\leq\epsilon''(C(K'))/\ell$, where $\epsilon''(C(K)):=\frac{C(K)^2}{\lambda_1^2(\Sigma_w)\,\lambda_1(Q)}\big(\lVert\Sigma_0\rVert+\frac{C(K)}{\lambda_1(Q)}\big)$ is increasing in $C$. Using \eqref{zero}, $\lVert U\rVert_F=v$ and $C(K+U)\leq C(K)+v\,h_C(C(K))$, we have $\lVert\Delta_D(K,v,\ell)\rVert_F\leq\bar\Delta_D(C(K),v,\ell)$ with 
\vspace{-5pt}
\begin{equation}\label{Deltak}
\bar\Delta_D(C(K),v,\ell):=\frac{n_xn_u\,\epsilon''\big(C(K)+vh_C(C(K))\big)}{v\,\ell}+v\,h_\nabla(C(K)),
\end{equation}
where $h_C,h_\nabla$ are the Lipschitz constants of $C$ and $\nabla C$ from Lemma~\ref{Lipschitz}. The $n$ rollouts are i.i.d., so $\mathbb{E}\lVert\hat\nabla_DC\rVert_F^2\leq\lVert\mathbb{E}\hat\nabla_DC\rVert_F^2+\frac{1}{n}\,\frac{n_x^2n_u^2}{v^4}\,\mathbb{E}\big[\hat C_{\bar K}^2\,\lVert U\rVert_F^2\big],$ where $\bar K=K+U$ and $\hat C_{\bar K}=\frac1\ell\sum_{t=0}^{\ell-1}x_t^\top Q_{\bar K}x_t$ with $Q_{\bar K}:=Q+\bar K^\top R\bar K$. Since $\lVert\mathbb{E}\hat\nabla_DC\rVert_F\leq b_\nabla(C(K))+\bar\Delta_D(C(K),v,\ell)$, the first term is at most $\phi(C(K),v,\ell):=\big(b_\nabla(C(K))+\bar\Delta_D(C(K),v,\ell)\big)^2 .$ For the second term, $\lVert U\rVert_F=v$, and by Cauchy--Schwarz $\hat C_{\bar K}^2\leq\frac1\ell\sum_{t}(x_t^\top Q_{\bar K}x_t)^2$. Since $x_0$ and $\{w_t\}$ are Gaussian and the closed loop is linear, each $x_t$ is zero-mean Gaussian given $U$ with covariance $\Sigma_{\bar K,t}:=\mathbb E[x_tx_t^\top\mid U]$, so $\mathbb{E}\big[(x_t^\top Q_{\bar K}x_t)^2\big|U]=\big(\mathrm{Tr}(Q_{\bar K}\Sigma_{\bar K,t})\big)^2+2\,\mathrm{Tr}\big((Q_{\bar K}\Sigma_{\bar K,t})^2\big)\leq3\big(\mathrm{Tr}(Q_{\bar K}\Sigma_{\bar K,t})\big)^2 .$ Moreover $\Sigma_{\bar K, t}\preceq A_{\bar K}^t\Sigma_0(A_{\bar K}^\top)^t+\Sigma_{\bar K}$ with $A_{\bar K}:=A+B\bar K$ and $\Sigma_{\bar K}$ the steady-state covariance driven by $\Sigma_w$, so $\mathrm{Tr}(Q_{\bar K}\Sigma_{\bar K})=C(\bar K)$ and $\mathrm{Tr}\big(Q_{\bar K}A_{\bar K}^t\Sigma_0(A_{\bar K}^\top)^t\big)\leq n_x\lVert\Sigma_0\rVert\,\frac{C(\bar K)}{\lambda_1(\Sigma_w)}$. Hence, for all $t$, $\mathrm{Tr}(Q_{\bar K}\Sigma_{\bar K, t})\leq c_{\hat C}\,C(\bar K)$ with $c_{\hat C}:=1+\frac{n_x\lVert\Sigma_0\rVert}{\lambda_1(\Sigma_w)}$, and with $C(\bar K)\leq C(K)+vh_C(C(K))$, $\mathbb{E}[\hat C_{\bar K}^2\mid U]\leq3\,c_{\hat C}^2\big(C(K)+v\,h_C(C(K))\big)^2 .$ Combining the above,
\vspace{-5pt}
\begin{equation}\label{Va2}
\begin{split}
        \mathbb{E}&\lVert\hat\nabla_DC(K,v,\ell,n)\rVert_F^2\\
        &\leq\phi(C(K),v,\ell)
    +\frac{3\,n_x^2n_u^2\,c_{\hat C}^2}{n\,v^{2}}\big(C(K)+v\,h_C(C(K))\big)^2=:V_D(C(K),v,\ell,n).
\end{split}
\end{equation}
Finally, \eqref{Deltak} gives $\bar\Delta_D=O\big(\frac{1}{v\ell}+v\big)$, and \eqref{Va2} gives $V_D=O\big(1+\frac{1}{\ell^2 v^2}+\frac{1}{nv^2}\big)$.
\end{proof}

\subsection{Proof of Theorem~\ref{thm:indirect}}\label{app:proof-indirect}
\begin{proof}
For Algorithm~\ref{alg:indirect}, the conditions of Theorems~\ref{thm:main-convergence} and~\ref{thm:RLS} hold only on the event to be established. We therefore analyze an auxiliary closed loop that, whenever one of these conditions fails, switches to the exact gradient or to zero and, in the on-policy case, freezes the excitation gain, so that it satisfies them surely, and that coincides with the algorithm on this event. A tilde marks its quantities; in particular, $\tilde F$, $\tilde G$ and $\widetilde{\mathrm{PE}}$ are the events $F$, $G$ and $\mathrm{PE}$ of Theorem~\ref{thm:indirect}, evaluated along the auxiliary loop defined later in \emph{Step 1}. Throughout, $p_\theta$, $p$ and $\bar g:=\bar g(J_0)$ are as in Lemma~\ref{lem:grad-error}, $c_x$ is as in \eqref{eq:RLS-mean}, $\Delta_0:=\Delta_0(J_0)$ is as in Theorem~\ref{thm:indirect}, and $\hat\nabla_IC(K,\theta')$ denotes \eqref{eq:indirect-grad} evaluated with the model $\theta'$.

\emph{Step 1 (auxiliary closed loop).} We define auxiliary sequences driven by the same initial state and noise sequences $\{w_t\}$, $\{e_t\}$ as Algorithm~\ref{alg:indirect}. The first $t_0$ samples are generated with $K_0$ as in Algorithm~\ref{alg:indirect}, so they coincide with the actual ones; set $\tilde K_0:=K_0$, $\tilde K^{\mathrm{ex}}_0:=K_0$, $\tilde H_0:=H_0$ and $\tilde\theta_0:=\hat\theta_0$. For $i\in\mathbb{Z}_+$, let $\widetilde{\mathrm{PE}}_i$ denote the event \eqref{eq:PE-events} for $\{\tilde H_k\}$, on which $\Delta\tilde\theta_i:=\tilde\theta_i-\theta$ is well defined, and let
\vspace{-5pt}
\begin{equation*}
    \tilde g_i:=\begin{cases}
    \hat\nabla_IC(\tilde K_i,\tilde\theta_i), & \text{if }\tilde K_i\in\mathcal{S}(J_0),\ \widetilde{\mathrm{PE}}_i\text{ holds, and }\|\Delta\tilde\theta_i\|_F\leq p_\theta,\\
    \nabla C(\tilde K_i), & \text{if }\tilde K_i\in\mathcal{S}(J_0)\text{ and one of the other two conditions fails},\\
    0, & \text{otherwise},
    \end{cases}
\end{equation*}
and $\tilde K_{i+1}:=\tilde K_i-\eta_i\tilde g_i$. In the off-policy case, $\tilde K^{\mathrm{ex}}_{i+1}:=K_0$. In the on-policy case, $\tilde K^{\mathrm{ex}}_{i+1}:=\tilde K_{i+1}$ if $\tilde K_k\in\mathcal{S}(J_0)$ for all $k\leq i+1$, and $\tilde K^{\mathrm{ex}}_{i+1}:=\tilde K^{\mathrm{ex}}_i$ otherwise. The sample at time $t_0+i$ is generated by $\tilde u_{t_0+i}=\tilde K^{\mathrm{ex}}_{i+1}\tilde x_{t_0+i}+e_{t_0+i}$, and $(\tilde\theta_{i+1},\tilde H_{i+1})$ is obtained from $(\tilde\theta_i,\tilde H_i)$ via \eqref{eq:RLS}. The auxiliary data are thus generated by $\bar K_t:=K_0$ for $t<t_0$ and $\bar K_{t_0+i}:=\tilde K^{\mathrm{ex}}_{i+1}$. Let $\mathcal{F}_i$ be the $\sigma$-algebra generated by the initial state and by $\{w_t,e_t\}_{t\leq t_0+i-1}$. Then $\widetilde{\mathrm{PE}}_i$, $\tilde\theta_i$, $\tilde g_i$, $\tilde K_{i+1}$ and $\tilde K^{\mathrm{ex}}_{i+1}$ are $\mathcal{F}_i$-measurable; in particular, $\bar K_t$ is independent of $\{(w_s,e_s)\}_{s\geq t}$. The auxiliary sequences are used only in the analysis: the conditions above depend on the unknown $\theta$ and are not observable.

\emph{Step 2 (mean-square boundedness).} We show that $\{\bar K_t\}$ satisfies, surely, the conditions of Lemma~\ref{lem:ms-bound}. Since $J_0>C(K_0)$, $K_0\in\mathcal{S}(J_0)$. In the off-policy case the gain is constant. In the on-policy case, $\tilde K^{\mathrm{ex}}_{i+1}$ is either $\tilde K^{\mathrm{ex}}_i$ or $\tilde K_{i+1}\in\mathcal{S}(J_0)$. Since the condition defining $\tilde K^{\mathrm{ex}}_{i+1}$ is decreasing in $i$, if it holds, then $\tilde K^{\mathrm{ex}}_i=\tilde K_i\in\mathcal{S}(J_0)$, so that $\|\tilde g_i\|_F\leq\bar g$ by Lemmas~\ref{bound} and~\ref{lem:grad-error}, and $\|\tilde K^{\mathrm{ex}}_{i+1}-\tilde K^{\mathrm{ex}}_i\|\le\eta_i\|\tilde g_i\|_F\le\bar\eta\,\bar g\le\varepsilon_K(J_0)$ by \eqref{eq:eta-variation}. Otherwise, $\tilde K^{\mathrm{ex}}_{i+1}=\tilde K^{\mathrm{ex}}_i$. Hence, by Theorem~\ref{thm:RLS} with \eqref{eq:t0-indirect}, $\vartheta=p_\theta$ and $\delta'=\delta_4$,
\vspace{-10pt}
\begin{equation}\label{eq:aux-RLS}
   \mathbb{E}\big[\|\Delta\tilde\theta_i\|_F\boldsymbol{1}_{\widetilde{\mathrm{PE}}_i}\big]\leq\frac{\sqrt{c_x}}{\alpha\sqrt{t_0+i}},\qquad
    \mathbb{P}(\tilde G\cap\widetilde{\mathrm{PE}})\geq1-\delta_4-\delta_5,
\end{equation}
where $\tilde G:=\{\|\Delta\tilde\theta_i\|_F\leq p_\theta,\ \forall i\in\mathbb{Z}_+\}$ and $\widetilde{\mathrm{PE}}:=\bigcap_i\widetilde{\mathrm{PE}}_i$.

\emph{Step 3 (the auxiliary oracle satisfies Assumption~\ref{as:oracle}).} The iteration $\tilde K_{i+1}=\tilde K_i-\eta_i\tilde g_i$ is of the form \eqref{eq:sgd-iteration} with filtration $\{\mathcal{F}_i\}$ and initial condition $\tilde K_0=K_0$, $C(K_0)<J_0$. Since $\tilde g_i$ is $\mathcal{F}_i$-measurable, $\mathbb{E}[\tilde g_i\mid\mathcal{F}_i]=\tilde g_i$, and the bias is $\tilde\Delta_i=\tilde g_i-\nabla C(\tilde K_i)$. On $\{\tilde K_i\in\mathcal{S}(J_0)\}$, Lemmas~\ref{bound} and~\ref{lem:grad-error} give $\|\tilde g_i\|_F\leq\bar g$ and $\|\tilde\Delta_i\|_F\leq p\,\|\Delta\tilde\theta_i\|_F\boldsymbol{1}_{\widetilde{\mathrm{PE}}_i}$, since the bias vanishes in the second case. Hence \eqref{eq:ABC} holds with $c(J_0)=\bar g^2$, and, by \eqref{eq:aux-RLS} and $t_0+i\geq i+1$, $\mathbb{E}\big[\|\tilde\Delta_i\|_F\boldsymbol{1}_{\{\tilde K_i\in\mathcal{S}(J_0)\}}\big]\leq p\,\mathbb{E}\big[\|\Delta\tilde\theta_i\|_F\boldsymbol{1}_{\widetilde{\mathrm{PE}}_i}\big]\leq\Delta_0\,(i+1)^{-1/2},$
i.e., \eqref{eq:bias-decay} holds with $\beta=\frac{1}{2}$. Since $C$ satisfies Definitions~\ref{def:GD}--\ref{def:local-smooth} and Assumption~\ref{as:grad-bound} by Lemmas~\ref{bound}--\ref{quasi}, Theorem~\ref{thm:main-convergence} applies to $\{\tilde K_i\}$. Let $\tilde F_i$ and $\tilde F:=\bigcap_i\tilde F_i$ denote the events \eqref{eq:event-Fi} and $F$ defined for $\{\tilde K_i\}$ in place of $\{X_i\}$, i.e., $\tilde K_k\in\mathcal{S}(J_0)$ for all $k\leq i$ and $\|\tilde K_k-\tilde K_{k-1}\|_F\leq r(J_0)$ for all $1\leq k\leq i$. Then $\mathbb{P}(\tilde F)\geq1-\delta$, and items 1--2 of Theorem~\ref{thm:main-convergence} hold for $\{\tilde K_i\}$ on $\tilde F$ with $\kappa>\beta=\frac{1}{2}$.

\emph{Step 4 (coupling).} Let $E:=\tilde F\cap\tilde G\cap\widetilde{\mathrm{PE}}$, so that $\mathbb{P}(E)\geq1-\delta-\delta_4-\delta_5$ by \eqref{eq:aux-RLS} and a union bound. We show by induction that, on $E$, $\tilde K_i=K_i$, $\tilde H_i=H_i$ and $\tilde\theta_i=\hat\theta_i$ for all $i\in\mathbb{Z}_+$. For $i=0$ this holds by construction. Suppose it holds up to $i$. On $E$, $\tilde K_i\in\mathcal{S}(J_0)$, $\widetilde{\mathrm{PE}}_i$ holds and $\|\Delta\tilde\theta_i\|_F\leq p_\theta$, so $\tilde g_i=\hat\nabla_IC(K_i,\hat\theta_i)=\hat g_i$ and $\tilde K_{i+1}=K_{i+1}$. Moreover, the condition defining $\tilde K^{\mathrm{ex}}_{i+1}$ holds on $E$, so $\tilde K^{\mathrm{ex}}_{i+1}=K^{\mathrm{ex}}_{i+1}$ in both cases. Since both loops are driven by the same noise and the same gains up to time $t_0+i$, they generate the same sample at that time, and $\tilde H_{i+1}=H_{i+1}$, $\tilde\theta_{i+1}=\hat\theta_{i+1}$. Hence, on $E$, $K_i=\tilde K_i$, $H_i=\tilde H_i$ and $\Delta\theta_i=\Delta\tilde\theta_i$ for all $i$, so that $E\subseteq\tilde F\cap\tilde G\cap\widetilde{\mathrm{PE}}$ yields $E\subseteq F\cap G\cap\mathrm{PE}$, and $\Delta_i=\tilde\Delta_i$ for all $i$. Since $E\subseteq\widetilde{\mathrm{PE}}_i$, \eqref{eq:aux-RLS} gives $\mathbb{E}\big[\|\Delta_i\|_F\boldsymbol{1}_E\big]\leq p\,\mathbb{E}\big[\|\Delta\theta_i\|_F\boldsymbol{1}_E\big]\leq p\,\mathbb{E}\big[\|\Delta\tilde\theta_i\|_F\boldsymbol{1}_{\widetilde{\mathrm{PE}}_i}\big]\leq\Delta_0(i+1)^{-1/2},$
which is \eqref{eq:indirect-bias-rate}. Similarly, $\mathbb{E}\big[(C(K_i)-C(K^*))\boldsymbol{1}_E\big]=\mathbb{E}\big[(C(\tilde K_i)-C(K^*))\boldsymbol{1}_E\big]\leq\mathbb{E}\big[(C(\tilde K_i)-C(K^*))\boldsymbol{1}_{\tilde F}\big],$
which yields item~1, and item~2 follows from $(C(K_i)-C(K^*))\boldsymbol{1}_E\leq(C(\tilde K_i)-C(K^*))\boldsymbol{1}_{\tilde F}$ and the almost-sure statement for $\{\tilde K_i\}$.
\end{proof}
\subsection{Proof of Theorem~\ref{thm:direct}}
\begin{proof}
Let $K\in\mathcal{S}(J_0)$. Since $v_i\leq v_0\leq h(J_0)\leq h(C(K))$ and $v_0\leq \tilde{c}$, Lemma~\ref{GDDI} applies with $(v,\ell,n)=(v_i,\ell_i,n_i)$. By \eqref{Deltak}, $\bar\Delta_D(C,v,\ell)=n_xn_u\,\epsilon''(C+vh_C(C))/(v\ell)\\+vh_\nabla(C)$, which is non-decreasing in $C$. Using $C(K)\leq J_0$, $v_i\leq v_0$, $v_i\ell_i\geq v_0\ell_0(i+1)^{1/2}$ and $v_i=v_0(i+1)^{-1/2}$, $\bar\Delta_D(C(K),v_i,\ell_i)\leq\Delta_0(J_0)(i+1)^{-1/2}.$
By \eqref{Va2}, $V_D(C,v,\ell,n)=\big(b_\nabla(C)+\bar\Delta_D(C,v,\ell)\big)^2+3n_x^2n_u^2c_{\hat C}^2\big(C+vh_C(C)\big)^2/(nv^2)$, which is non-decreasing in $C$. Since $\bar\Delta_D(J_0,v_i,\ell_i)\leq\Delta_0(J_0)$ and $n_iv_i^2\geq n_0v_0^2$, we obtain $\bar V_D(J_0)<\infty$, which proves \eqref{temp4}. Since $K_i$ is $\mathcal{F}_i$-measurable and the perturbations, initial states, and noise of the rollouts at iteration $i$ are independent of $\mathcal{F}_i$, Lemma~\ref{GDDI} holds conditionally on $\mathcal{F}_i$ with $K=K_i$. Hence, on $\{K_i\in\mathcal{S}(J_0)\}$, $\mathbb{E}[\hat g_i\mid\mathcal{F}_i]=\nabla C(K_i)+\Delta_D(K_i,v_i,\ell_i)$, $\mathbb{E}\big[\|\hat g_i\|_F^2\mid\mathcal{F}_i\big]\leq\bar V_D(J_0)$, 
and $\|\Delta_D(K_i,v_i,\ell_i)\|_F\leq\Delta_0(J_0)(i+1)^{-1/2}$ by \eqref{temp4}. In particular, \eqref{eq:ABC} holds with $c(J_0)=\bar V_D(J_0)$, and \eqref{eq:bias-decay} holds with $\beta=\frac{1}{2}$, since the bias is bounded deterministically on $\{K_i\in\mathcal{S}(J_0)\}$.
Since $C$ satisfies Definitions~\ref{def:GD}--\ref{def:local-smooth} and Assumption~\ref{as:grad-bound}, Theorem~\ref{thm:main-convergence} applies with $f=C$, $X_i=K_i$, $c(J_0)=\bar V_D(J_0)$, $\Delta_0(J_0)$ and $\beta=\frac{1}{2}$, which yields $\mathbb{P}(F)\geq1-\delta$ and items 1--2 with $\kappa>\beta=\frac{1}{2}$. 
\end{proof}
\subsection{Full expressions}\label{DetailedExpression}
\begin{subequations}
\begin{align}
    b_{\nabla}(C(K))&:=\frac{2C(K)b_E(C(K))}{\lambda_1(Q)}\label{boundedgradienteq};\\
    b_{K}(C(K))&:= \frac{\lVert B \rVert \lVert A \rVert C(K)}{\lambda_1(\Sigma_w)\lambda_1(R) }+ \frac{b_E(C(K))}{\lambda_1(R) \label{boundK}};\\
    b_E(C(K))&:=\sqrt{ \frac{(C(K)-C(K^*))}{\textcolor{black}{\lambda_1(\Sigma_w)}}\left(  \lVert R \rVert+ \frac{\lVert B \rVert^2 C(K)}{{\lambda_1{(\Sigma_w)}}}\right)}.\label{boundedEK}
\end{align}
\end{subequations}
\vspace{-15pt}
\begin{subequations}
    \begin{align}
    \begin{split}\label{Pertubation1}
        h_{C}(C(K)) &:= 2\left( \frac{C(K)}{\lambda_1(\Sigma_w)\lambda_1(Q)}\right)^2\lVert R \rVert\big( b^2_{K}(C(K))\lVert B \rVert(\lVert A\rVert\\&\quad+\lVert B \rVert b_{K}(C(K))+1)+2b_{K}(C(K))+\tilde{c}\big)\mathrm{Tr}(\Sigma_w);
    \end{split}\\
    \begin{split}\label{ErrorGradient}
        h_{\nabla}(C(K))&:= 2h_{\Sigma}{(C(K))}b_E(C(K))+2\lVert R \rVert+ \frac{\lVert B \rVert^2 C(K)}{\lambda_1{(\Sigma_w)}}\\&\quad+\frac{h_C(C(K))}{\mathrm{Tr}(\Sigma_w)}\left(  \lVert B \rVert\lVert A \rVert+(b_{K}(C(K))+\tilde{c}) \lVert B \rVert^2 \right).
    \end{split}
    \end{align}
\end{subequations}

\bibliographystyle{siamplain}
\bibliography{referencenew271026}
\end{document}